\documentclass[12pt]{iopart}
\usepackage{iopams}
\usepackage{graphicx}
\expandafter\let\csname equation*\endcsname\relax
\expandafter\let\csname endequation*\endcsname\relax
\usepackage{amsmath}
\usepackage{amssymb}
\usepackage{cite}

\begin{document}

\title[First return times of non-backtracking random walks]
{Analytical results for the distribution of  
first return times of non-backtracking random walks on 
configuration model networks
}

\author{Dor Lev-Ari$^1$, Ido Tishby$^1$, Ofer Biham$^{1,*}$, 
Eytan Katzav$^1$ and Diego Krapf$^2$}
\address{$^1$ Racah Institute of Physics, 
The Hebrew University, Jerusalem 9190401, Israel}
\address{$^2$ Department of Electrical and Computer Engineering 
and School of Biomedical Engineering, Colorado State University, 
Fort Collins, Colorado 80523, USA}

\eads{\mailto{dor.lev-ari@mail.huji.ac.il},
\mailto{ido.tishby@mail.huji.ac.il}, 
\mailto{ofer.biham@mail.huji.ac.il}, 
\mailto{eytan.katzav@mail.huji.ac.il},
\mailto{diego.krapf@colostate.edu}
}

\begin{abstract}

We present analytical results for the distribution of
first return (FR) times of
non-backtracking random walks (NBWs) on 
undirected configuration model networks
consisting of $N$ nodes with degree distribution $P(k)$.
We focus on the case in which the network consists 
of a single connected component.
Starting from a random initial node $i$ at time $t=0$, 
an NBW hops into a random neighbor of $i$ at time $t=1$
and at each subsequent step it continues to hop
into a random neighbor of its current node,
excluding the previous node.
We calculate the tail distribution 
$P ( T_{\rm FR} > t )$
of first return times from a random initial node to itself.
It is found that
$P ( T_{\rm FR} > t )$
is given by a discrete Laplace transform of the degree distribution $P(k)$.
This result exemplifies the relation between structural properties of a network,
captured by the degree distribution,
and properties of dynamical processes taking place on the network.
Using the tail-sum formula, we calculate the mean first return time
${\mathbb E}[ T_{\rm FR} ]$. 
Surprisingly,
${\mathbb E}[ T_{\rm FR} ]$
coincides with the result obtained from Kac's lemma that applies to simple 
random walks (RWs).
We also calculate the variance ${\rm Var}(T_{\rm FR})$, which accounts
for the variability of first return times between different NBW trajectories.
We apply this formalism to 
Erd{\H o}s-R\'enyi networks,
random regular graphs
and configuration model networks with exponential and
power-law degree distributions and obtain closed-form expressions for
$P( T_{\rm FR} > t )$ as well as its mean and variance.
These results provide useful insight on the advantages of NBWs over
simple RWs in network exploration, sampling and search processes.

\end{abstract}


\noindent
$^*$ Author to whom any correspondence should be addressed. 

\maketitle

\section{Introduction}

Random walk (RW) models
\cite{Lawler2010}
provide useful tools for the analysis of
dynamical processes on random networks 
\cite{Havlin2010,Newman2010,Barrat2012,Masuda2017,Hofstad2017,Hofstad2024}.
Here we focus on the case of undirected networks.
Starting at time $t=0$ from a random initial node $i$, 
at each time step $t \ge 1$ an RW
(also referred to as a simple RW)
hops randomly to one of the neighbors of its current node.
In some of the time steps the RW visits nodes that
have not been visited before, while
in other time steps it revisits nodes that have
already been visited at an earlier time.
The mean number $\langle S \rangle_t$ of distinct nodes visited by an RW 
on a random network up to time $t$ was studied in Ref.
\cite{Debacco2015}.
It was found that 
in the infinite network limit, 
in which random networks exhibit a tree structure,
$\langle S \rangle_t \simeq r t$, 
where the coefficient
$r<1$ depends on the network topology.
In this case, the revisits are due to backtracking steps in which the RW hops back to the 
previous node and subsequent
retroceding steps in which it keeps hopping backwards along
its own path
\cite{Tishby2021}.

In order to perform systematic studies of random walks on random networks, 
it is useful to focus on configuration model networks.
The configuration model is an ensemble 
of uncorrelated random networks
consisting of $N$ nodes,
whose degree sequences are drawn from
a given degree distribution $P(k)$.
The admissible degrees are often restricted to a finite range
$k_{\rm min} \le k \le k_{\rm max}$,
where $k_{\rm min}$ is the minimal degree and $k_{\rm max}$ is the maximal degree,
such that for any value of $k$ outside this range $P(k)=0$.
The mean degree $\langle K \rangle$ is denoted by $c$.
To ensure that an RW starting from any initial node $i$ will be able to 
reach any other node $j$, we focus on the case in which the whole
network consists of a single connected component.
Using the terminology of percolation theory, these are networks in which
the giant component encompasses the whole network.
In the large network limit,
a sufficient condition for a configuration model network to consist of a
single connected component is that $k_{\rm min} \ge 3$
\cite{Bollobas2001,Hofstad2017}.
In fact, a weaker condition of $k_{\rm min} \ge 2$ is sufficient
in the large network limit as long as a finite fraction of the
nodes satisfy $k \ge 3$
\cite{Federico2023}.
We thus avoid isolated nodes of degree $k=0$
and leaf nodes of degree $k=1$, 
which may form isolated tree structures.

The first return (FR) time $T_{\rm FR}$ of an RW  
is the first time at which it returns to the initial node $i$
\cite{Redner2001}.
The first return time varies between different instances of the random walk
trajectory and its properties can be captured by a suitable distribution.
The distribution of first return times may depend on the 
specific realization of the random network and on the
choice of the initial node $i$.
The distribution of first return times 
from a random node to itself
in a given ensemble of
random networks is denoted by $P(T_{\rm FR}=t)$.
A classical result regarding first return times is Kac's lemma,
which states that the mean first return time 
of an RW 
from a given node $i$ to itself
is given by
\cite{Kac1947,Levin2009,Dorogovtsev2022}

\begin{equation}
{\mathbb E}[T_{\rm FR}(i)] = \frac{1}{P_i(\infty)},
\label{eq:Kac1}
\end{equation}

\noindent
where $P_i(\infty)$ is the probability that an RW
will reside at node $i$ at a given time step
under steady state conditions 
(which are achieved in the long time limit $t \rightarrow \infty$).
In the case of undirected networks, Eq. (\ref{eq:Kac1})
can be expressed in a more explicit form, namely
\cite{Dorogovtsev2022}

\begin{equation}
{\mathbb E}[T_{\rm FR}(i)] = \frac{Nc}{k_i},
\label{eq:Kac2}
\end{equation}

\noindent
where $k_i$ is the degree of node $i$.
This implies that the mean first return time from a random node 
to itself is given by

\begin{equation}
{\mathbb E}[T_{\rm FR}] = \bigg\langle \frac{Nc}{K} \bigg\rangle,
\label{eq:Kac3}
\end{equation}

\noindent
where $\langle X \rangle$ 
is the average of 
the random variable
$X$ over the degree distribution $P(k)$.

One can distinguish between two types of first return trajectories:
first return trajectories in which the RW
retrocedes its own steps backwards all the way back to the initial node $i$
and first return trajectories in which the RW returns to $i$ 
via a path that does not retrocede its own steps
\cite{Tishby2021,Tishby2022b}.
In the retroceding trajectories, each edge that belongs to the RW trajectory is crossed the
same number of times in the forward and backward directions.
In the non-retroceding trajectories the RW path
includes at least one cycle.
In the infinite system limit, in which the network exhibits a tree structure,
the only way to return to the initial node is via a retroceding trajectory
\cite{Masuda2004,Evnin2024}.
In finite networks both scenarios coexist, where 
the distribution $P(T_{\rm FR}=t)$ is dominated by
retroceding trajectories  
at short times and by non-retroceding trajectories at long times
\cite{Tishby2021,Hormann2024,Albert2024}.

A more general problem involves the calculation of
the first passage (FP) time $T_{\rm FP}$,
which is the first time at which a random walk starting from an initial node $i$
at time $t=0$
visits a specified target node $j$
\cite{Redner2001,Sood2005,Peng2021,Tishby2022b}.
The first return problem is a special case of the first passage problem, 
in which the initial node coincides with the target node.
The distribution $P(T_{\rm FR}=t)$ of first return times of RWs was studied on the
Bethe lattice, which exhibits a tree structure of an infinite size
\cite{Hughes1982,Cassi1989,Giacometti1995,Masuda2004,Martin2010}
and on random regular graphs (RRGs)
\cite{Tishby2021,Hormann2024}.

An important variant of the RW model is the non-backtracking random walk (NBW),
in which the move backwards to the previous node is excluded
\cite{Fitzner2013}.
Since backtracking steps are excluded, in the infinite network limit in which
the network exhibits a tree structure, an NBW never revisits a previously
visited node.
In particular, it never returns to the initial node.
In a finite network, the first return process of NBWs takes place 
only via non-retroceding trajectories,
which rely on the existence of cycles.

NBWs are important for several reasons, which are summarised below.
They provide a more efficient way to explore and analyze complex networks,
compared to standard random walks. 
This is due to the fact that by avoiding backtracking steps 
they can cover more of the network in less time
\cite{Alon2007}.
NBWs are useful for identifying community structures within networks
\cite{Arrigo2018}. 
In certain types of networks, NBWs can help mitigate localization effects that might trap 
standard random walks in specific regions of the network.

In this paper we present analytical results for 
the distribution of first return times 
of NBWs on configuration model networks consisting of $N$ nodes 
with degree distribution $P(k)$.
An NBW starting from an initial node $i$ forms a random trajectory in
the network and eventually returns to $i$ without backtracking its steps even once.
In order to return to the initial node, the trajectory must include at least one cycle.
The first return time may take any even or odd
value that satisfies $T_{\rm FR} \ge 3$.
Using probabilistic methods we calculate the tail-distribution of first return times
$P(T_{\rm FR}>t|K=k)$
of NBWs starting from a random node of degree $k$.
Averaging over the degree distribution, 
we obtain the overall tail distribution of first return times
$P(T_{\rm FR}>t)$.
We find that
$P(T_{\rm FR}>t)$
is given by a discrete Laplace transform of the degree distribution $P(k)$.
We calculate the mean first return time ${\mathbb E}[ T_{\rm FR} ]$
and show that it coincides with the result of Fasino et al. 
\cite{Fasino2023},
which extends Kac's lemma to second order random walks.
We also calculate the variance ${\rm Var}(T_{\rm FR})$ 
which accounts for the variability of first return times
between different NBW trajectories.
We apply this formalism to random regular graphs, Erd{\H o}s-R\'enyi
networks and configuration model networks with exponential and
power-law degree distributions
and obtain closed-form expressions for
$P ( T_{\rm FR} > t )$ and its first two moments.
The analytical results are found to be in very good agreement with the results 
obtained from computer simulations.

The paper is organized as follows.
In Sec. 2 we present the configuration model networks, 
their construction and essential properties.
In Sec. 3 we present the non-backtracking random walk.
In Sec. 4 we derive formulae for the distribution $P(T_{\rm FR}>t)$ 
of first return times of NBWs on configuration 
model networks and for its mean ${\mathbb E}[T_{\rm FR}]$ 
and variance ${\rm Var}(T_{\rm FR})$.
In Sec. 5 we apply these results to RRGs,
Erd{\H o}s-R\'enyi (ER) networks and to configuration model networks with
exponential and power-law distributions.
The results are discussed in Sec. 6 and summarised in Sec. 7.

\section{Configuration model networks}

The configuration model is an ensemble 
of uncorrelated random networks
whose degree sequences are drawn from
a given degree distribution $P(k)$
\cite{Bollobas1980,Molloy1995,Molloy1998,Newman2001,Fosdick2018}.
These  networks are simple graphs in the sense that
each pair of nodes is connected by at most a single edge
and there are no self-loops.
The first moment (mean degree) and the second moment 
of $P(k)$ are denoted by
$\langle K^n \rangle$,
where $n=1$ and $2$, respectively,
while the variance is given by
$Var(K) = \langle K^2 \rangle - \langle K \rangle^2$.
The support of the degree distribution
of random networks is often bounded from below
by $k_{\rm min} \ge 1$ such that $P(k)=0$
for $0 \le k \le k_{\rm min}-1$,
with non-zero values of
$P(k)$ only for
$k \ge k_{\rm min}$.
For example, the commonly used choice of $k_{\rm min}=1$
eliminates the possibility of isolated nodes in the network. 
Choosing $k_{\rm min}=2$ also eliminates the leaf nodes.
One may also control the upper bound by imposing
$k \le k_{\rm max}$.
This may be important in the case of finite networks
with heavy-tail degree distributions  
such as power-law distributions.
The configuration model network ensemble is a maximum entropy ensemble
under the condition that the degree distribution $P(k)$ is imposed
\cite{Molloy1995,Molloy1998,Newman2001}.
Here we focus on the case of undirected networks.

To generate a network instance drawn from an ensemble of
configuration model networks of $N$ nodes,
with a given degree distribution $P(k)$, one draws
the degrees of the $N$ nodes independently from 
$P(k)$.
This gives rise to a
degree sequence of the form
$k_1,k_2,\dots,k_N$.
For the discussion below it is convenient to list the degree
sequence in a decreasing order of the form
$k_1 \ge k_2 \ge \dots \ge k_N$.
It turns out that not every possible degree sequence is graphical,
namely admissible as a degree sequence of a network.
Therefore, before trying to construct a network with a given
degree sequence, one should first confirm
the graphicality of the degree sequence.
To be graphical, a degree sequence must satisfy two conditions.
The first condition is that the sum of the degrees is an even number,
namely
$\sum_{i} k_i = 2 L$,
where $L$ is an integer that represents
the number of edges in the network.
The second condition is expressed by the Erd{\H o}s-Gallai theorem,
which states that an ordered sequence of the form
$k_1 \ge k_2 \ge \dots \ge k_N$
that satisfies the first condition
is graphical if and only if the condition
\begin{equation}
\sum_{i=1}^n k_i \le n(n-1) + \sum_{i=n+1}^N \min (k_i,n)
\label{eq:EG}
\end{equation}

\noindent
holds for all values of $n$ in the range
$1 \le n \le N-1$
\cite{Erdos1960b,Choudum1986}.

To construct a network instance consisting of $N$ nodes 
with a given degree sequence $k_1, k_2,\dots,k_N$
(where $\sum_{i=1}^{N} k_i = 2L$ and $L$ is the total number of undirected edges),
we create a multiset of $2L$ stubs which includes $k_i$ stubs for each node $i$.
Pairs of stubs are then selected randomly and connected to each other
to form edges between the corresponding nodes.
To illustrate the process we represent the stubs by $2L$ balls, where the $k_i$ balls
associated with node $i$ are marked by $i$.
We then choose a random arrangement of the $2L$ balls
in an array of $L$ cells, such that each cell includes exactly two balls.
In practice, a random arrangement of balls into cells can be obtained by generating 
a random permutation of the $2L$ balls and grouping them sequentially into $L$
pairs, making the construction straightforward to implement.
A cell containing balls 
$i$ and $j$ represents an edge between nodes $i$ and $j$.
The representation in terms of balls and cells is particularly convenient for implementation
on the computer, since a single random permutation of the $2L$ balls produces a 
uniformly random pairing of stubs, from which the network can be constructed directly.

The network obtained from the procedure described above is a multigraph
with the given degree sequence,
which may include self-loops
(edges connecting a node to itself)
or multiple edges (two or more edges connecting the same pair of nodes).
To eliminate the self-loops and multiple edges, 
we apply an  
edge switching process, 
which yields a simple graph while preserving the degree sequence.
In this process, 
as long as the network has not yet become a simple graph,
at each time step we select randomly one of the self-loops $(i,i)$ or one of the multiple edges
$(i,j)$. 
In case that a self-loop $(i,i)$ was selected, we select
a random edge $(i',j')$ and swap the two edges into $(i,i')$ and $(i,j')$.
Similarly, in case that a multiple edge $(i,j)$ was selected, we
select a random edge $(i',j')$ and swap the two edges into $(i,i')$ and $(j,j')$.
In both cases, we complete the move only 
after we make sure that the swapping does not create a new self-loop or a new multiple edge. 
This random edge-switching process continues until no self-loops or multiple edges remain. 
The procedure described above provides the random simple graph ensemble used in the simulations.

The elimination of multiple edges may introduce some degree-degree correlations
in the resulting simple graph.
To keep these degree-degree correlations negligible, 
the degree distribution must exhibit a structural cutoff such that
the expected number of nodes of degree $k > \sqrt{ N c }$ 
is $o(1)$
\cite{Boguna2004,Catanzaro2005}.
In the case of fat-tailed degree distributions such as the 
power-law degree distribution, one needs to impose an upper cutoff
$k_{\rm max} < \sqrt{ N c }$.
The degree distributions of all the network models considered here satisfy the above conditions,
so degree-degree correlations are negligible.

Some commonly studied configuration model networks can be 
described in terms of single parameter families of degree distributions.
These include the RRG, the ER network and configuration model
networks with exponential and power-law degree distributions.
A particularly convenient choice of the parameter is the mean degree
$c=\langle K \rangle$. 
In this case, the degree distribution can be expressed by
$P(k)=P_c(k)$, such that small values of $c$ correspond to
the dilute network limit while large values of $c$ correspond to the
dense network limit.

Configuration model networks in which the lower bound of the degree distribution satisfies
$k_{\rm min}=0$ or $1$, may exhibit a percolation transition at 
some value $c_0$ of the mean degree, referred to as the percolation threshold.
Below the transition the network consists of finite tree components,
while above the transition a giant component emerges.
The percolation transition is a second order phase transition, 
whose order parameter is the fraction $g$ of nodes that reside on
the giant component.
Below the transition, where $c < c_0$, the order parameter is $g=0$,
while for $c > c_0$ the fraction $g=g(c)$ 
of nodes that reside on the giant component gradually increases.
The giant component of a configuration model network consists of a $2$-core
which is decorated by tree branches
\cite{Tishby2018b}.
The $2$-core is a connected component, such that each
node on the $2$-core has links to at least two other nodes that
reside on the $2$-core.
The nodes that reside on the tree branches have the property that
their deletion would break the giant component into two or more
components.
Such nodes are referred to as articulation points
\cite{Tian2017,Tishby2018}.
Similarly, the deletion of an edge that resides on one of the tree branches
would break the giant component into two components.
Such edges are referred to as bredges
\cite{Bonneau2020}.

In this paper we focus on the case in which the whole network
consists of a single connected component, for which $g=1$.
Below we discuss the conditions for $g=1$ in RRGs, ER networks 
and configuration model networks with exponential 
and power-law distributions.

Consider an RRG that consists of $N$ nodes of degree $c$
(where $Nc$ is even).
For $c=1$ the nodes form dimers.
For $c=2$ the network,
which is referred to as a $2$-random regular graph ($2$-RRG),
consists of closed loops or cycles.
In the large network limit, the expected number of cycles is 
$N_C \simeq \frac{1}{2} \ln N$
and the cumulative distribution of cycle lengths is given by
$P(L \le \ell) \simeq \ln \ell/\ln N$,
where $\ell \le N$
\cite{Tishby2023}.
Here we focus on RRGs with $c \ge 3$,
which in the large network limit consist
of a single connected component
\cite{Bender1978,Wormald1981}.

In the case of ER networks, in the large network limit
there is a phase transition at $c_1 = \ln N$, 
where $g \rightarrow 1$
\cite{Bollobas2001,Hofstad2017}. 
Above this point, the giant component encompasses the whole network,
linking all the nodes into a single connected component.

In general, a sufficient condition for a configuration model
network with degree distribution $P(k)$ to consist of a single
connected component in the large network limit 
$N \rightarrow \infty$
is $k_{\rm min} \ge 3$
\cite{Wormald1999}.
In fact, a weaker condition of $k_{\rm min} \ge 2$ is also sufficient,
as long as a finite fraction of the nodes in the network are of degrees
$k \ge 3$
\cite{Federico2023}.
In the analysis presented below of NBWs on configuration model
networks with exponential and power-law distributions, we chose
networks of size $N=1000$ that satisfy $k_{\rm min} \ge 3$. 
We checked each network instance to confirm that it consists
of a single connected component.

In a finite configuration model network, 
there is a non-zero probability that the network 
will consist of a single connected component 
even if it includes some nodes of degree 
$k=1$.
It was recently shown 
\cite{Federico2017}
that as the network size $N$ is increased,
it may still consist of a single connected component with high probability
as long as the number $n_1$ of nodes of degree $k=1$ grows
more slowly than $\sqrt{N}$.
However, in networks that include leaf nodes of degree $k=1$, 
NBWs that enter these nodes will get stuck. 
Therefore, in the study of NBWs it is important not only to
ensure that the network consists of a single connected component, 
but also that this component does not include any leaf nodes.
This implies that the $2$-core of the network 
(namely the largest subgraph in which all the nodes are of degree $k \ge 2$)
encompasses the whole network.
It also implies that the network does not include any articulation points
\cite{Tian2017,Tishby2018}
or bredges
\cite{Bonneau2020}.

\section{Non-backtracking random walks}

NBWs are RWs for which the move backwards to the previous node is excluded.
They belong to the class of second-order random walks, in which the
transition probabilities depend not only on the current node but also on the
previous node
\cite{Fitzner2013,Fasino2023}.
This introduces memory into the process, 
which makes it no longer Markovian in the traditional sense.
The challenge is to analyze such processes using 
methods that are typically applied to Markov chains,
which rely on the memoryless property.
Recently, Fasino et al. introduced a mapping of second order 
random walks into first order processes
on a larger state space, referred to as the pullback process
\cite{Fasino2023}.
Instead of viewing the random walk as taking place 
between the nodes of the original graph,
the pullback process considers a random walk on the 
directed-line graph associated with the original graph.
Using this method they showed that the mean first return time 
${\mathbb E}[T_{\rm FR}]$
of any second order random walk
(including NBWs) on undirected networks
satisfies Eq. (\ref{eq:Kac1}), thus extending the validity of Kac's 
lemma to second order random walks on undirected networks
\cite{Fasino2023}.
Note that Kac's lemma deals with the mean first return 
time and has no implications
on the overall shape of the distribution and its higher order moments.

NBWs exhibit faster mixing times than standard random walks, 
meaning they converge to their stationary distribution more quickly
\cite{Cioba2015,Hamou2019}.
They thus inspire the design of more efficient 
algorithms for various graph-based problems, 
including link prediction and node centrality measures
\cite{Wu2016}.
The non-backtracking (Hashimoto)
matrix $B$ associated with these walks has spectral properties 
that can reveal important information about the network structure, 
often more clearly than traditional adjacency matrices
\cite{Krzakala2013,Karrer2014,Dorogovtsev2022}.
Actually, the mixing time is inversely proportional 
to the spectral gap of the matrix $B$
\cite{Levin2009}

\begin{equation}
t_{\rm mix} \propto \frac{1}{ \vert \lambda_1 - \lambda_2 \vert },
\end{equation}

\noindent
where $\lambda_1$ and $\lambda_2$ are the largest and 
second largest eigenvalues of $B$,
respectively.

For the special case of RRGs it was shown that 
the mixing time of NBWs (and RWs)
scales like $t_{\rm mix} \propto \ln N$
\cite{Alon2007,Lubetzky2010}.
This result was later generalized to a broader class of configuration
model networks with $k_{\rm min} \ge 3$
\cite{Hamou2017}.
This result sits well with the fact that both the mean distance 
\cite{Hofstad2005,Tishby2022}
and the diameter 
\cite{Bollobas1982,Fernholz2007,Riordan2010,Chung2001,Mieghem2023}
of RRGs are proportional to $\ln N$.
It implies that an NBW starting from a random initial node $i$ at time $t=0$
may reach any other node in the network within $\ln N$ time steps.
Moreover, using the shell structure around the initial node $i$ as  
a spherical coordinate system, 
the radial component of the location of each node is given by its 
distance from $i$.
Since RRGs are locally tree-like at distances in the range $\ell \ll \ln N$
\cite{Bonneau2017},
an NBW starting from $i$ essentially moves deterministically to the
next shell away from $i$ as far as the tree-like structure persists.
This is unlike the case of RWs which behave like biased random walks
along the radial axis, moving outwards with probability $1 - 1/c$ and 
inwards with probability $1/c$
\cite{Tishby2021,Tishby2022}.

\section{The distribution of first return times}

Consider an NBW on an undirected random network, 
starting from a random initial node $i$ at time $t=0$. 
At time $t=1$ it hops into a random neighbor of $i$ and
at each subsequent step it
hops randomly into one of the neighbors of its current node,
excluding the previous node.
Here we focus on the case of
configuration model networks that
consist of a single connected component,
such that an NBW starting from any initial node
can reach any other node in the network.

At each time step $t \ge 3$ an NBW may either step into a yet-unvisited node or
into a node that has already been visited two or more time steps earlier.
Similarly, at each time step $t \ge 4$ an NBW may go through an edge 
from node $i$ to node $i'$, that has been
crossed before in the same direction, or through an 
edge that has not yet been crossed 
in that direction.
We thus distinguish between the two possibilities of
crossing an edge: from $i$ to $i'$ and 
from $i'$ to $i$.
In a network of size $N$ and mean degree $c$,
the expected number of such 'directed' edges is $Nc$.
Below we consider the expected number of distinct 'directed' edges 
$\langle L \rangle_t$ crossed by an NBW up to time $t$
on a configuration model network.
The initial condition is $\langle L \rangle_0 = 0$.
The probability that at time step $t$ an NBW will cross a yet 
uncrossed 'directed' edge is given by

\begin{equation}
\Delta L_t =  \langle L \rangle_{t+1} - \langle L \rangle_{t}.
\label{eq:DeltaL1}
\end{equation}

In the first three time steps the NBW crosses new `directed' edges with probability $1$, 
which implies that $\Delta L_t = 1$
for $t = 0, 1$ and $2$.
For $t \ge 3$ we use a mean-field approach,
which essentially assumes that the `directed' edges that have already been 
crossed and those that have not yet been crossed are distributed uniformly
in the network and can thus be visited with equal probability at any time step.
This approach applies under the condition that the network consists of a single
connected component.
A further condition is that the network will not be dominated by linear chains
consisting of nodes of degree $k=2$, namely that $P(k=2)$ will be sufficiently small
(networks that do not satisfy this condition are referred to as almost $2$-RRGs
\cite{Federico2023}).

In configuration model networks that consist of a single
connected component with $k_{\rm min} \ge 3$ and no leaf nodes,
the mixing time scales like $t_{\rm mix} \propto \ln N$
\cite{Lubetzky2010,Hamou2017},
while the mean first return time scales like 
${\mathbb E}[T_{\rm FR}] \propto N$.
Thus, for sufficiently large networks
$t_{\rm mix} \ll  \mathbb{E}[T_{\rm FR}]$.
This separation of time scales implies that apart
from the very early stages of the first return trajectories,
NBWs sample the `directed' edges in a
uniform fashion.
Under these conditions, the probability that at time $t+1$ the NBW will cross
a `directed' edge which has been crossed before is equal to the
fraction of `directed' edges that have already been crossed.
This fraction is given by $(\langle L \rangle_t - 2)/(Nc - 2)$,
where the subtraction of $2$ from the numerator and the denominator
accounts for the fact that the `directed' edges crossed at times $t-1$ and $t$ cannot
be crossed again at time $t+1$.
This implies that the probability $\Delta L_t$ 
is given by

\begin{equation}
\Delta L_t = 1 - \frac{\langle L \rangle_t - 2}{Nc - 2}.
\label{eq:DeltaL2f}
\end{equation}

\noindent
To simplify the analysis, we reduce Eq. (\ref{eq:DeltaL2f})
to the form

\begin{equation}
\Delta L_t = 1 - \frac{\langle L \rangle_t}{Nc}.
\label{eq:DeltaL2}
\end{equation}

\noindent
The reduction from Eq. (\ref{eq:DeltaL2f}) to Eq. (\ref{eq:DeltaL2}) relies on the assumption
that the network is both sufficiently large and sufficiently dense, such that
the product $Nc$ 
satisfies $N c \gg 2$.
In addition, this reduction becomes accurate when
the expected number of `directed' edges $\langle L \rangle_t$
which have already been visited by the NBW satisfies $\langle L \rangle_t \gg 1$.
This condition is indeed satisfied for sufficiently long times.
Since for $t \ll Nc$ the number of distinct `directed' edges
visited by the NBW satisfies $\langle L \rangle_t \simeq t$,
the condition 
$\langle L \rangle_t \gg 1$
can be replaced by $t \gg 1$.

Inserting $\Delta L_t$ from Eq. (\ref{eq:DeltaL1}) into Eq. (\ref{eq:DeltaL2}), 
we obtain the recursion equation  

\begin{equation}
\langle L \rangle_{t+1} =
 \langle L \rangle_{t}
\left(1 - \frac{ 1 }{Nc}  \right) + 1.
\label{eq:L_t_RE}
\end{equation}

\noindent
Solving Eq. (\ref{eq:L_t_RE}), 
we obtain

\begin{equation}
\langle L \rangle_t = 
\left\{
\begin{array}{ll}
t   & \ \ \ \ \ \   0 \le t \le 3   \\
3 e^{ - \frac{t-3}{Nc} } + Nc \left( 1 - e^{ - \frac{t-3}{Nc} } \right)    & \ \ \ \ \ \  t > 3.  \\
\end{array}
\right.
\label{eq:L_t1}
\end{equation}

\noindent
While Eq. (\ref{eq:DeltaL2}) is valid to a good approximation for $t \ge 3$,
it becomes precise above the mixing time,
where the random walker samples `directed' edges in a uniform fashion.
Thus, apart from the first few steps,
Eq. (\ref{eq:L_t1}) can be approximated by

\begin{equation}
\langle L \rangle_t = 
  Nc \left( 1 - e^{ - \frac{t}{Nc} } \right).  
\label{eq:L_t2}
\end{equation}

The probability that an NBW will not visit a specific
random node of degree $k$ for the first time up to time $t$,
can be expressed by $\left( 1 - \frac{\langle L \rangle_{t-1}}{Nc} \right)^k$.
This is due to the fact that in order to visit a node of degree $k$ the NBW must
enter via one of the $k$ edges connected to $i$.
Since an NBW quickly loses memory of its initial node, 
the probability of not returning to an initial node of degree $k$ up to time $t$
is the same as the probability not to visit any node of the
same degree up to time $t$. 
Therefore, the tail distribution of first return times, 
under the condition that the initial node $i$ is
of degree $k$, is given by

\begin{equation}
P(T_{\rm FR} > t | K=k) = \left( 1 - \frac{\langle L \rangle_{t-1}}{Nc} \right)^k.
\label{eq:PTFRk}
\end{equation}

\noindent
Inserting $\langle L \rangle_{t-1}$ from Eq. (\ref{eq:L_t1}) into Eq. (\ref{eq:PTFRk}), 
and using the fact that $k \ll Nc$, we obtain the
tail distribution of first return times for initial nodes of degree $k$,
which is given by

\begin{equation}
P(T_{\rm FR} > t | K=k) = 
e^{ - \frac{k}{Nc} t  }.  
\label{eq:PTFRk2}
\end{equation}

To obtain the tail distribution 
$P(T_{\rm FR} > t)$
of first return times of an NBW starting from a random
node, we average over all possible initial nodes. 
This amounts to averaging over all
possible degrees, with weights given by $P(k)$. We obtain

\begin{equation}
P(T_{\rm FR} > t) = \sum_{k=0}^{\infty}
e^{ - \frac{t}{Nc} k }
P(k).
\label{eq:PTFRk3}
\end{equation}

\noindent
Interestingly, the right hand side of Eq. (\ref{eq:PTFRk3})
is a discrete Laplace transform of the degree distribution $P(k)$.
This transform is related to the one-sided Z-transform and to the starred transform
\cite{Phillips2015}.
To illustrate this point, we express Eq. 
(\ref{eq:PTFRk3})
in the form

\begin{equation}
P(T_{\rm FR} > t) = \sum_{k=0}^{\infty}
z^k P(k),
\label{eq:PTFRk4}
\end{equation}

\noindent
where

\begin{equation}
z = e^{ - \frac{t}{Nc}  }.
\end{equation}

\noindent
In fact, the right hand side of Eq. (\ref{eq:PTFRk4}) 
is equal to the generating function $G_0(z)$ 
of the degree distribution $P(k)$.
The generating function is known to play a central role in the analysis of
structural properties of random networks such as the percolation 
threshold 
\cite{Newman2001}
and the distribution of shortest path lengths
\cite{Nitzan2016}.
Therefore, Eq. (\ref{eq:PTFRk4}) provides a remarkable connection between structural
properties of a network, captured by $G_0(z)$ and properties of dynamical 
processes taking place on the network.

From known properties of the (discrete) Laplace transform, we infer that
the tail of $P(T_{\rm FR}>t)$ is determined by the abundances of the lowest degree
nodes at the left end of $P(k)$. 
In contrast, the left end of $P(T_{\rm FR}>t)$ is determined by
the highest degree nodes (or hubs) in the tail of $P(k)$.

The probability mass function of first return times is given by the difference

\begin{equation}
P(T_{\rm FR}=t) = P(T_{\rm FR}>t-1) - P(T_{\rm FR}>t).
\end{equation}

\noindent
The moments of the distribution of first return times
can be obtained from the 
tail-sum formula
\cite{Pitman1993}.
In particular, the mean first return time is given by

\begin{equation}
{\mathbb E}[T_{\rm FR}] = \sum_{t=0}^{\infty} P(T_{\rm FR}>t),
\label{eq:ETFR0}
\end{equation}

\noindent
and the second moment is given by

\begin{equation}
{\mathbb E} \left[ T_{\rm FR}^2 \right] = \sum_{t=0}^{\infty} 
(2t+1) P(T_{\rm FR}>t).
\label{eq:ET2FR0}
\end{equation}

\noindent
The variance 
is given by

\begin{equation}
{\rm Var}(T_{\rm FR}) = 
{\mathbb E} \left[ T_{\rm FR}^2 \right]
- 
{\mathbb E}[T_{\rm FR}]^2.
\label{eq:VarTFR}
\end{equation}
 
To evaluate the mean first return time, we insert $P(T_{\rm FR}>t)$
from Eq. (\ref{eq:PTFRk3}) into Eq. (\ref{eq:ETFR0}) and obtain

\begin{equation}
{\mathbb E}[T_{\rm FR}]
= 
\sum_{t=0}^{\infty}  
 \sum_{k=0}^{\infty}
e^{ - \frac{k}{Nc} t }
P(k).
\label{eq:ETFR1}
\end{equation}

\noindent
Exchanging the order of the summations and carrying out the sum over $t$,
we obtain

\begin{equation}
{\mathbb E}[T_{\rm FR}]
= 
\sum_{k=0}^{\infty}
\frac{1}{1 - e^{ - \frac{k}{Nc} } }
P(k).
\label{eq:ETFR2}
\end{equation}

\noindent
Expanding the exponent in the denominator in terms of $k/(Nc) \ll 1$ and taking the
leading term, we obtain

\begin{equation}
{\mathbb E}[T_{\rm FR}]
\simeq
\sum_{k=0}^{\infty}
\frac{Nc}{k}
P(k) = \bigg\langle \frac{Nc}{K} \bigg\rangle.
\label{eq:ETFR3}
\end{equation}

\noindent
This result coincides with Kac's lemma,
which is obtained from general properties of discrete stochastic processes
\cite{Kac1947,Fasino2023}.
Eq. (\ref{eq:ETFR3}) also implies that 
conditioning on initial nodes of a given degree $k$, the mean first return
time is given by

\begin{equation}
{\mathbb E}[T_{\rm FR} | K=k]
\simeq
\frac{Nc}{k}.
\label{eq:ETFR4}
\end{equation}

Eq. (\ref{eq:ETFR3}) implies that the mean first return time
is proportional to the mean inverse degree $\big\langle \frac{1}{K} \big\rangle$.
In order to express this quantity in terms of $\langle K \rangle$ and ${\rm Var}(K)$,
one can use a Taylor expansion of $1/K$ around $\langle K \rangle$ and obtain
its expectation value

\begin{equation}
\bigg\langle \frac{1}{K} \bigg\rangle =
\frac{1}{ \langle K \rangle }
+ \frac{1} { \langle K \rangle^3 } {\rm Var} (K)
- \frac{1} { \langle K \rangle^4 }
\big\langle \left( K - \langle K \rangle \right)^3 \big\rangle  + \dots.
\label{eq:inverseK}
\end{equation}

\noindent
This expansion is suitable for narrow distributions 
that are concentrated around their mean value.
Moreover, in the case of symmetric distributions the 
third term on the right hand side of
Eq. (\ref{eq:inverseK}) vanishes and the first two terms 
are expected to provide accurate results
for $\big\langle \frac{1}{K} \big\rangle$.
Within the domain of validity of Eq. (\ref{eq:inverseK}), 
we conclude that the mean first return time
${\mathbb E}[T_{\rm FR}]$
of an NBW on a configuration model network
is proportional to the variance of the degree distribution
of the network.

Using a similar derivation for the second moment, 
which is based on Eq. (\ref{eq:ET2FR0}),
we obtain

\begin{equation}
{\mathbb E} \left[T_{\rm FR}^2 \right]
= 
\sum_{k=0}^{\infty}
\frac{1 + e^{ - \frac{k}{Nc} } }{\left( 1 - e^{ - \frac{k}{Nc} } \right)^2 }
P(k)
\simeq
2 \sum_{k=0}^{\infty}
\left( \frac{Nc}{k} \right)^2 P(k) 
=
2 \bigg\langle \left( \frac{Nc}{K} \right)^2 \bigg\rangle.
\label{eq:ET2FR}
\end{equation}

\noindent
This result goes beyond the generalization of Kac's 
lemma for second-order random walks
\cite{Fasino2023}
and is valid for the specific case of the NBW.

Conditioning on initial nodes of a given degree $k$, 
the second moment is given by

\begin{equation}
{\mathbb E} \left[T_{\rm FR}^2 | K=k \right]
\simeq
2 \left( \frac{Nc}{k} \right)^2  
\label{eq:ET2FRp}
\end{equation}

\noindent
Inserting 
${\mathbb E} \left[T_{\rm FR}^2 \right]$
from Eq. (\ref{eq:ET2FR})
and
${\mathbb E}[T_{\rm FR}]$
from Eq. (\ref{eq:ETFR3})
into Eq. (\ref{eq:VarTFR}),
we obtain

\begin{equation}
{\rm Var}(T_{\rm FR}) 
\simeq
N^2 c^2
\left(
2 \bigg\langle \frac{1}{K^2} \bigg\rangle
- 
\bigg\langle \frac{1}{K} \bigg\rangle^2
\right). 
\label{eq:VarTFR2}
\end{equation}

\noindent
Thus, the variance of the distribution of first return times, 
conditioned on initial nodes of a given degree $k$,
is given by

\begin{equation}
{\rm Var}(T_{\rm FR} | K=k) = \frac{N^2 c^2}{k^2}.
\label{eq:VarTFk}
\end{equation}

\noindent
This indicates that the variance ${\rm Var}(T_{\rm FR})$ 
of the distribution of first return times
can be divided into two parts,  according to

\begin{equation}
{\rm Var}(T_{\rm FR}) = 
N^2 c^2 \  {\rm Var} \left( \frac{1}{K} \right)
+ 
N^2 c^2
\bigg\langle \frac{1}{K^2} \bigg\rangle,
\label{eq:VarTFR3}
\end{equation}

\noindent
where the first term on the right hand side of Eq. (\ref{eq:VarTFR3})
can be attributed to the variations in the degrees between different 
initial nodes, while the second term can be attributed to the variation
in the first return times between NBW trajectories originating from 
nodes of the same degree.

\section{Application to specific random networks}

In this section we apply the general results derived above to
NBWs on specific random networks that belong 
to the class of configuration model networks
\cite{Newman2001}.
In particular we study the first return process on
RRGs, ER networks and configuration model networks with exponential
and power-law degree distributions.
For each type of network we calculate the tail distribution of first return times
as well as the mean and variance.

\subsection{Random regular graphs}

Random regular graphs are random networks of a finite size 
in which all the nodes are of the same degree,
but the connectivity is random
\cite{Bollobas2001}. 
They thus belong to the class of configuration model networks.
Consider an RRG that consists of $N$ nodes of degree $c \ge 3$.
In such network, 
in the large $N$ limit,
all the nodes reside on a single connected component.
As a result, an RW (or NBW) starting from any initial 
node $i$ may reach any other node $j$.

The degree distribution of an RRG is a degenerate distribution of the form

\begin{equation}
P(k)=\delta_{k,c},
\label{eq:Pkdeg}
\end{equation}

\noindent
where the mean degree $\langle K \rangle = c$ is an integer and the variance 
${\rm Var}(K) = 0$.

Inserting $P(k)$ from Eq. (\ref{eq:Pkdeg})
into Eq. (\ref{eq:PTFRk3}), we obtain the
tail distribution of first return times, which is given by

\begin{equation}
P(T_{\rm FR} > t) = e^{- \frac{t}{N}}.
\label{eq:PTFR_RRG}
\end{equation}

It would be useful to compare the distribution 
of first return times of NBWs on RRGs
to the corresponding distribution of simple RWs on RRGs.
The latter distribution consists of a contribution from
retroceding trajectories, which are dominant at short times
and non-retroceding trajectories, which are dominant at long times.
The distribution $P(T_{\rm FR} > t)$, given by Eq. (\ref{eq:PTFR_RRG}),
is analogous to the contribution of the non-retroceding RW trajectories
in simple RWs,
which for sufficiently long times is given by
\cite{Tishby2021}

\begin{equation}
P(T_{\rm FR} > t | \lnot {\rm RETRO}   ) = 
\exp \left[ -    \left( \frac{c-2}{c-1} \right)   \frac{t}{N}     \right] .    
\label{eq:P_FRT_tail2}
\end{equation}

\noindent
This implies that the backtracking and retroceding 
steps slow down the first return process
of RWs by a factor of $\frac{c-2}{c-1}$ compared to NBWs.

\begin{figure}
\centerline{
\includegraphics[width=9cm]{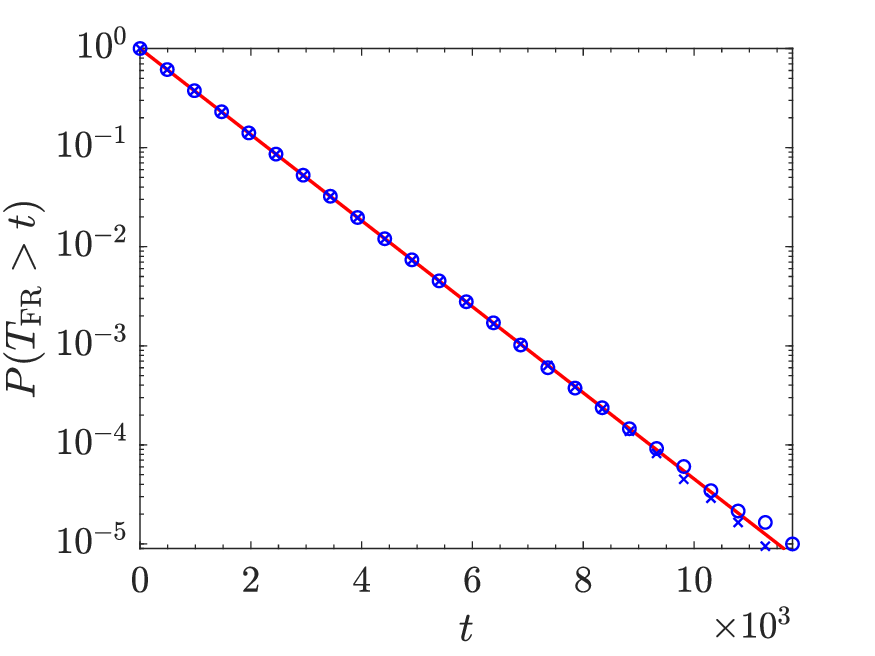}
}
\caption{
Analytical results (solid line),
obtained from Eq. (\ref{eq:PTFR_RRG}),
for the 
tail distribution  
$P(T_{\rm FR} > t)$ 
of first return times
of an NBW on an RRG
of size $N=1000$.
The right hand side of Eq. (\ref{eq:PTFR_RRG})
does not depend on the degree $c$,
which implies that these results are valid for RRGs
with any degree $c \ge 3$.
Indeed, the analytical results
are in very good agreement with the results
obtained from computer simulations
for $c=3$ ($\times$) and for $c=10$ ($\circ$).
Each data point of the simulation results was obtained by
averaging the results obtained for $20$ network instances 
and $100,000$ NBW trajectories for each network instance.
}
\label{fig:1}
\end{figure}

In Fig. \ref{fig:1} we present
analytical results, 
obtained from Eq. (\ref{eq:PTFR_RRG}),
for the 
tail distribution  
$P(T_{\rm FR} > t)$ (solid line)
of first return times
of an NBW on an RRG
of size $N=1000$.
Note that the right hand side of Eq. (\ref{eq:PTFR_RRG})
does not depend on the degree $c$,
which implies that the results are valid for RRGs
with any degree $c \ge 3$.
Indeed, the analytical results
are found to be in very good agreement with the results
obtained from computer simulations
for RRGs with $c=3$ ($\times$) and $c=10$ ($\circ$).

The mixing time of an NBW on an RRG of size $N$ and degree $c$
is given by 
\cite{Lubetzky2010}

\begin{equation}
t_{\rm mix}(N,c) = \frac{ \ln N }{ \ln (c-1) } + {\mathcal O}(1).
\end{equation}

\noindent
Applying this result to NBWs on RRGs of size $N=1000$ with degrees 
of $c=3$ and $c=10$,
it is found that
$t_{\rm mix}(1000,3) \simeq 10$
and
$t_{\rm mix}(1000,10) \simeq 3$,
which are clearly much smaller than 
the time scales that are relevant to the first return process.

For the simulations we generated $20$ independent instances
of the network. On each network instance, we generated $100,000$ NBW
trajectories, where each trajectory starts from a random initial node $i$ at time $t=0$.
Each NBW trajectory was terminated upon its first return
to the initial node $i$. The first return time $t$ is thus equal to the length of the trajectory. 
The simulation results were obtained by averaging the results over all these trajectories.

Inserting Eq. (\ref{eq:PTFR_RRG}) into Eq. (\ref{eq:ETFR3}),
we obtain the mean first return time,
which is given by

\begin{equation}
{\mathbb E}[T_{\rm FR}]  \simeq N,
\label{eq:ETFRrrg}
\end{equation}

\noindent
thus the mean first return time does not depend on the degree $c$.
This result is in agreement with Kac's lemma, expressed by Eq. (\ref{eq:Kac1}).
Since all the nodes in an RRG are of the same degree, the probability that an RW
(or an NBW) will reside at any given node at time $t$ is 
$P_i(\infty) = 1/N$. Inserting $P_i(\infty)$ into Eq. (\ref{eq:Kac1}), we obtain
${\mathbb E}[T_{\rm FR}] = N$.
This result is also in agreement with the mean first return time of a simple RW on an RRG,
calculated in Ref. 
\cite{Tishby2021}.

Similarly, one can calculate the second moment, which is given  by

\begin{equation}
{\mathbb E} \left[ T_{\rm FR}^2  \right]  \simeq 2N^2.
\end{equation}

\noindent
Therefore, the variance is

\begin{equation}
{\rm Var}(T_{\rm FR})  \simeq N^2.
\label{eq:VarFRrrg}
\end{equation}

\noindent
Since in an RRG all the nodes are of the same degree, this variance reflects the variability
between first return trajectories originated from nodes of the same degree.
Going back to Eq. (\ref{eq:VarTFR3}), we conclude that Eq. (\ref{eq:VarFRrrg}) 
represents the lowest possible variance in the distribution of first return times
for random networks consisting of $N$ nodes.

Interestingly, for a simple RW on an RRG, it was found that 
the variance of the distribution of first return times is given by
\cite{Tishby2021}

\begin{equation}
{\rm Var}(T_{\rm FR})  \simeq  \frac{c}{c-2} N^2.
\end{equation}

\noindent
This result is larger than the variance for NBWs by a multiplicative factor of $\frac{c}{c-2}$.
This factor is significant for sparse RRGs and approaches $1$ as $c$ is increased.
It is due to the fact that in simple RWs the distribution of first return times is bimodal,
consisting of two different types of first return trajectories.
At short times it is dominated by retroceding trajectories while at long times
it is dominated by non-retroceding trajectories.
This separation of time scales broadens the distribution and increases the variance.
The difference in the variance between NBWs and simple RWs
reflects the fact that Kac's
lemma applies only to the mean first return time and does not 
provide any prediction for the variance.

\subsection{Erd{\H o}s-R\'enyi networks}

Consider an Erd{\H o}s-R\'enyi network that consists of $N$ nodes. 
In such network, each pair of nodes is connected by an edge with
probability $p$
\cite{Erdos1959,Erdos1960,Erdos1961}.
As a result, the degree distribution is a Poisson 
distribution of the form
\cite{Newman2010}

\begin{equation}
P(k) = \frac{ e^{-c} c^k }{k!},
\label{eq:PkER}
\end{equation}

\noindent
for $k=0,1,2,\dots$,
where $c=(N-1)p$ is the mean degree $\langle K \rangle$
and the variance is given by 
${\rm Var}(K) = c$.

In general, for $c > 1$ an ER network consists of a giant component
and finite tree components. 
Since we focus in this paper on networks that consist of a single connected
component, we restrict ourselves to the case in which $c > \ln N$,
where in the large network limit
the giant component encompasses the whole network
\cite{Bollobas2001,Hofstad2017}.
In the case that $c > \ln N$, the probability that a random 
node will be isolated is $P(K=0) < 1/N$,
which implies that in a typical network instance the 
expected number of isolated nodes will be smaller than $1$.
Since we study NBWs we would like to ensure that the 
network instances we consider will also
not include leaf nodes of degree $k=1$. 
Therefore, in the analysis we focus on the limit
of sufficiently dense networks that satisfy 
$c > - W(-1/N)$, where $W(x)$ is the Lambert W function
\cite{Olver2010}. In this limit the probability that a 
random node will be a leaf node satisfies
$P(K=1) < 1/N$.
In practice, when we generate network instances 
for the computer simulations, we discard
network instances that include isolated nodes or leaf nodes.

For an NBW starting from a random node $i$ on an ER network,
the tail distribution of first return times is obtained by inserting
Eq. (\ref{eq:PkER}) into Eq. (\ref{eq:PTFRk3}), which yields

\begin{equation}
P(T_{\rm FR}>t) = \exp \left[ c \left( e^{- \frac{t}{Nc}} - 1 \right) \right].
\label{eq:PTFR_ER1}
\end{equation}

\noindent
Note that in the long time limit of $t \rightarrow \infty$,
$P(T_{\rm FR}>t) \rightarrow e^{-c}$,
which is bounded by $1/N$ for $c > \ln N$
and hence vanishes in the large system limit.
However, for finite networks the fact that
$P(T_{\rm FR}>t)$ does not vanish in the limit
of $t \rightarrow \infty$ and therefore the moments diverge. 
In order to deal with this issue, we adjust the degree distribution
by eliminating the possibility of isolated nodes of
degree $k=0$ and leaf nodes of degree $k=1$. 
The adjusted degree distribution is
given  by

\begin{equation}
P(k|K>1) = \frac{1}{ 1 - e^{-c} - c e^{-c} } \frac{ e^{-c} c^k }{k!},
\label{eq:PkER0}
\end{equation}
 
\noindent
for $k \ge 2$.
Inserting the adjusted degree distribution
from Eq. (\ref{eq:PkER0}) into Eq. (\ref{eq:PTFRk3}),
we obtain

\begin{equation}
P(T_{\rm FR}>t|K>1) = \frac{e^{-c}}{ 1-e^{-c} - c e^{-c} }    
\left[  \exp \left( c   e^{- \frac{t}{Nc}}     \right) - 1 - c e^{- \frac{t}{Nc}  } \right].
\label{eq:PTFR_ER2}
\end{equation}

\noindent
Taking the long time limit of Eq. (\ref{eq:PTFR_ER2}), 
we obtain the leading order asymptotic
behavior, which exhibits an exponential tail of the form

\begin{equation}
P(T_{\rm FR}>t|K>1) \simeq \frac{c^2 e^{-c}}{2(1-e^{-c}-c e^{-c})} e^{ - \frac{2}{Nc} t }.
\end{equation}

\noindent
This tail is dominated by the lowest degree nodes 
in the network, whose degree is $k=2$.

\begin{figure}
\centerline{
\includegraphics[width=9cm]{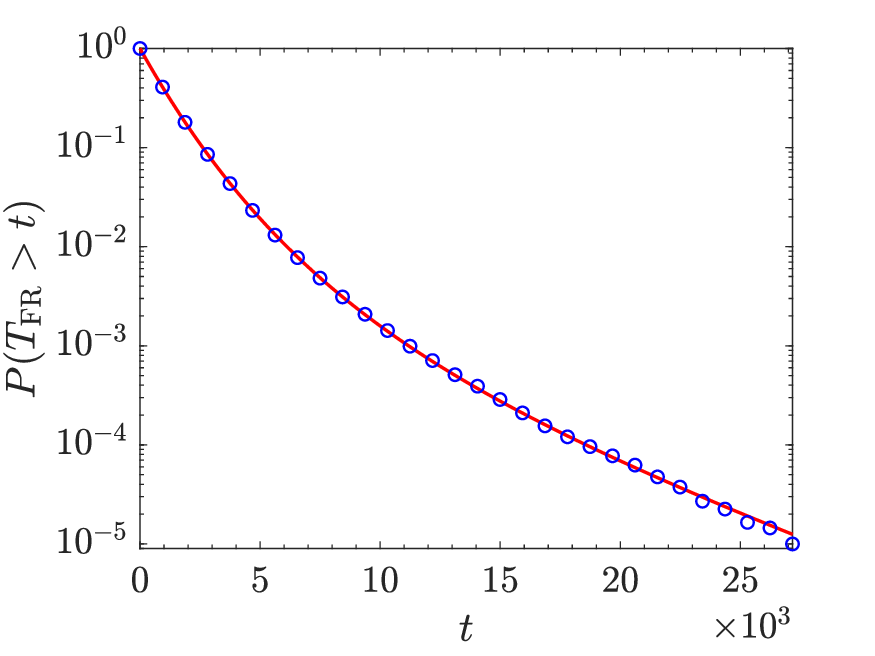}
}
\caption{
Analytical results for the 
tail distribution  
$P(T_{\rm FR} > t | K>1)$ (solid line)
of first return times
of an NBW on an Erd{\H o}s-R\'enyi network
of size $N=1000$ 
and mean degree 
$c=10$.
The analytical results,
obtained from Eq.
(\ref{eq:PTFR_ER2}),
are in very good agreement with the results
obtained from computer simulations (circles).
The simulation results were obtained using the same
averaging procedure as in Fig. \ref{fig:1}.
}
\label{fig:2}
\end{figure}

In Fig. \ref{fig:2} we present
analytical results for the 
tail distribution  
$P(T_{\rm FR} > t | K>1)$ (solid lines)
of first return times
of an NBW on an Erd{\H o}s-R\'enyi network
of size $N=1000$ 
and mean degree 
$c=10$.
The analytical results,
obtained from Eq.
(\ref{eq:PTFR_ER2}),
are in very good agreement with the results
obtained from computer simulations (circles).

Inserting 
$P(T_{\rm FR}>t|K>1)$
from Eq. (\ref{eq:PTFR_ER2}) into Eq. (\ref{eq:ETFR3}),
we obtain

\begin{equation}
{\mathbb E}[ T_{\rm FR} | K>1] = 
{\mathbb E} \left[ \frac{Nc}{K} \bigg\vert K>1 \right].
\label{eq:ETFRk03}
\end{equation}

\noindent
Evaluating the mean on the right hand side of Eq. (\ref{eq:ETFRk03}), we obtain

\begin{equation}
{\mathbb E}[ T_{\rm FR} | K>1] = 
Nc \frac{e^{-c}}{ 1 - e^{-c} - c e^{-c} }
\left[ {\rm Ei}(c) - c - \ln c - \gamma \right], 
\label{eq:ETFRk04}
\end{equation}

\noindent
where ${\rm Ei}(x)$ is the exponential integral 
\cite{Olver2010}

\begin{equation}
Ei(x) = \int_{- \infty}^{x} \frac{e^t}{t} dt,
\label{eq:Ei}
\end{equation}

\noindent
and $\gamma$
is the Euler-Mascheroni constant
\cite{Olver2010}.
In the limit of large mean degree $c$,
Eq. (\ref{eq:ETFRk04}) can be simplified to

\begin{equation}
{\mathbb E}[ T_{\rm FR} | K>1]  = 
N \left[
1 + \frac{1}{c} + {\mathcal O} \left( \frac{1}{c^2} \right)
\right],
\label{eq:ETFRk05}
\end{equation}

\noindent
where ${\mathcal O}(1/c^2)$ means that the terms of order $1/c^2$ and higher
are ignored in the expansion.
This is in agreement with the first two terms on the right hand side of 
Eq. (\ref{eq:inverseK}), confirming the validity of the expansion to the 
Poisson distribution, for sufficiently large values of the mean degree $c$.
Eq. (\ref{eq:ETFRk05}) shows that the mean first return time in an ER network 
is larger than in an RRG of the same size, and is a decreasing function of $c$.

\begin{figure}
\centerline{
\includegraphics[width=9cm]{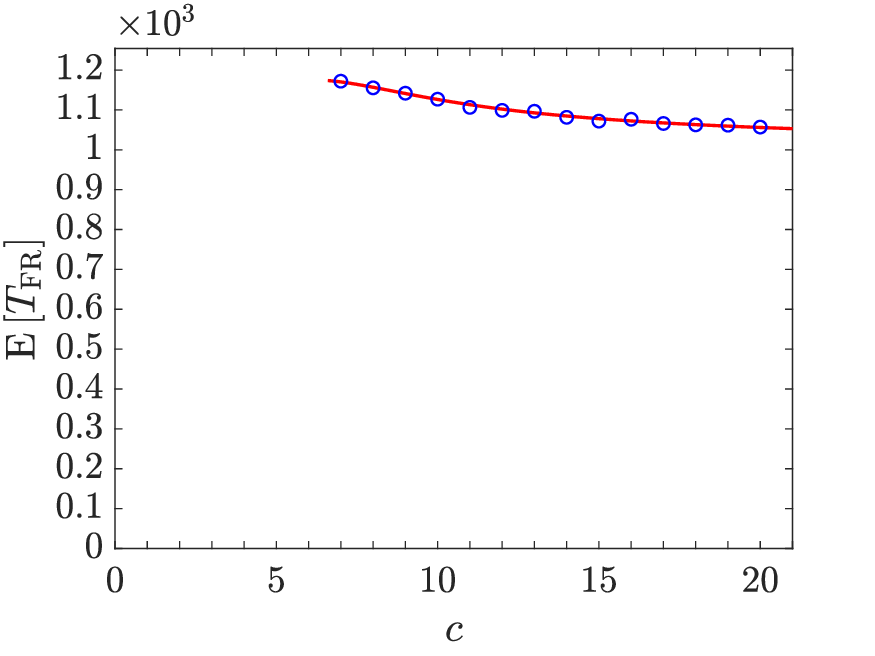}
}
\caption{
Analytical results for the
mean first return time 
${\mathbb E}[ T_{\rm FR} | K>1 ]$ (solid line)
of an NBW on an Erd{\H o}s-R\'enyi network of size $N=1000$,
as a function of the mean degree 
$c$, for $c > \ln N$, where the whole network consists of a single connected component
and network instances that include leaf nodes are discarded.
The analytical results,
obtained from Eq.
(\ref{eq:ETFRk04}), 
are in very good agreement with 
the results obtained from computer simulations (circles).
Each data point of the simulation results was obtained by
averaging the results obtained for $20$ network instances 
and $10,000$ NBW trajectories for each network instance.
}
\label{fig:3}
\end{figure}

In Fig. \ref{fig:3} we present
analytical results for the
mean first return time 
${\mathbb E} [ T_{\rm FR} | K>1 ]$ (solid line)
of an NBW on an Erd{\H o}s-R\'enyi network of size $N=1000$,
as a function of the mean degree 
$c$.
The analytical results,
obtained from Eq.
(\ref{eq:ETFRk04}), 
are in very good agreement with 
the results obtained from computer simulations (circles).

Similarly, we can calculate the second moment, by plugging Eq. (\ref{eq:PkER0})
into Eq. (\ref{eq:ET2FR}).
We obtain

\begin{equation}
{\mathbb E} \left[ T_{\rm FR}^2 | K>1 \right] \simeq
2 N^2 c^2 \frac{ e^{-c} }{ 1 - e^{-c} - c e^{-c} }
\sum_{k=2}^{\infty} 
\frac{c^k}{k!} \frac{1}{k^2}. 
\label{eq:ET2k0ER3}
\end{equation}

\noindent
Carrying out the summation on the right hand side of Eq. (\ref{eq:ET2k0ER3}), we obtain

\begin{equation}
{\mathbb E} \left[ T_{\rm FR}^2 | K>1 \right] \simeq
2 N^2 c^3 \frac{ e^{-c} }{ 1 - e^{-c} - c e^{-c} }   
\left[  \, _3 F_3
\left(\left. \begin{array}{c}
1,1,1 \\
2,2,2 
\end{array}
\right|
c \right) - 1 \right].
\label{eq:ET2k0ER4}
\end{equation}

\noindent
where 
$\, _3 F_3
\Bigl( \Big. \begin{array}{c}
a_1,a_2,a_3 \\
b_1,b_2,b_3
\end{array}
  \Big|
z \Bigr)$
is the generalized hypergeometric function
\cite{Olver2010}.
Thus, the variance of $P(T_{\rm FR} > t | K > 1)$ is given by

\begin{eqnarray}
{\rm Var} (T_{\rm FR} | K>1) &=&
2 N^2 c^3 \frac{ e^{-c} }{1 - e^{-c} - c e^{-c} }
\left[
\, _3 F_3
\left(\left. \begin{array}{c}
1,1,1 \\
2,2,2 
\end{array}
\right|
c \right)
- 1
\right]
\nonumber \\
&-&
N^2 c^2 \left( \frac{e^{-c}}{1-e^{-c} - c e^{-c} } \right)^2
\left[ {\rm Ei}(c) - c - \ln c - \gamma \right]^2.
\label{eq:VarTFRK0_ER}
\end{eqnarray}

\noindent
In the limit of large mean degree $c$,
one can simplify Eq. (\ref{eq:VarTFRK0_ER}),
which takes the form

\begin{equation}
{\rm Var} (T_{\rm FR} | K>1) =
N^2 \left[ 1 + \frac{4}{c}  + {\mathcal O} \left( \frac{1}{c^2} \right) \right].
\label{eq:VarTFRK0_ER2}
\end{equation}

\begin{figure}
\centerline{
\includegraphics[width=9cm]{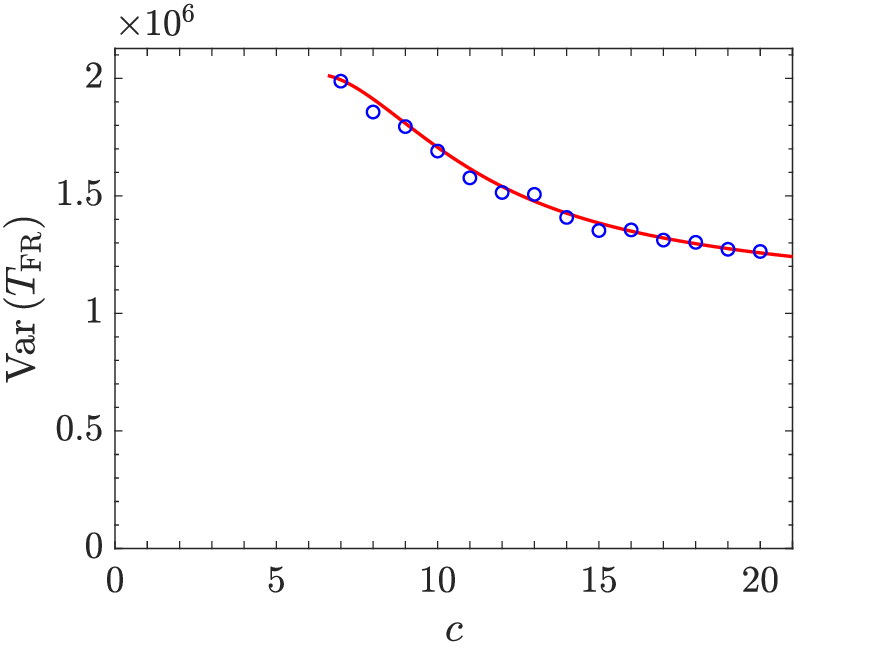}
}
\caption{
Analytical results for the
variance   
${\rm Var}(T_{\rm FR} | K>1)$ (solid line)
of the distribution of first return times
of an NBW on an Erd{\H o}s-R\'enyi network of size $N=1000$,
as a function of the mean degree 
$c$.
The analytical results,
obtained from Eq.
(\ref{eq:VarTFRK0_ER}), 
are in very good agreement with 
the results obtained from computer simulations (circles).
The simulation results were obtained using the same averaging procedure as in Fig. \ref{fig:3}.
}
\label{fig:4}
\end{figure}

In Fig. \ref{fig:4} we present 
analytical results for the
variance   
${\rm Var}(T_{\rm FR} | K>1)$ (solid line)
of the distribution of first return times
of an NBW on an Erd{\H o}s-R\'enyi network of size $N=1000$,
as a function of the mean degree 
$c$.
The analytical results,
obtained from Eq.
(\ref{eq:VarTFRK0_ER}), 
are in very good agreement with 
the results obtained from computer simulations (circles).

\subsection{Configuration model networks  with an exponential degree distribution}

Consider an ensemble of configuration model networks with an exponential degree
distribution of the form 

\begin{equation}
P(k) = A e^{- \alpha k}, 
\label{eq:Exponential}
\end{equation}

\noindent
where $\alpha > 0$ is the rate parameter and the degree $k$ takes values in the range
$k_{\rm min} \le k \le \infty$
[$P(k)=0$ for $0 \le k \le k_{\rm min} - 1$].
The parameter $A$ is a normalization factor and it is given by
$A = (1-e^{-\alpha}) e^{ \alpha k_{\rm min} }$.
In order to obtain a network that consists of a single connected component,
one needs to choose $k_{\rm min} \ge 2$.

Imposing the normalization condition and parameterizing the distribution in terms
of the mean degree $c = \langle K \rangle$, 
one can rewrite the degree distribution in the form
\cite{Tishby2018}

\begin{equation}
P(k) = \frac{1}{c - k_{\rm min} + 1}
\left( \frac{ c - k_{\rm min} }{c - k_{\rm min} + 1} \right)^{k - k_{\rm min}},
\label{eq:Pk_Exp1}
\end {equation}

\noindent
for $k \ge k_{\rm min}$.
The parameter $\alpha$ from Eq. (\ref{eq:Exponential}) can be expressed in the form

\begin{equation}
\alpha = \ln \left( \frac{ c - k_{\rm min} + 1}{c - k_{\rm min}} \right).
\label{eq:c_exp}
\end{equation}

\noindent
The variance of the exponential degree distribution can be expressed in the form

\begin{equation}
{\rm Var}(K) = (c - k_{\rm min} + 1)(c - k_{\rm min}),
\label{eq:VarKexp}
\end{equation}

\noindent
such that in the limit of a broad degree distribution,
where $c \gg k_{\rm min}$ it can be approximated by

\begin{equation}
{\rm Var}(K) \simeq  c^2.
\label{eq:VarKexp2}
\end{equation}

Inserting $P(k)$ from Eq. (\ref{eq:Pk_Exp1}) into Eq. (\ref{eq:PTFRk3}) and carrying out the summation,
we obtain the distribution of first return times

\begin{equation}
P(T_{\rm FR} > t) =
\frac{1}{c+1-k_{\rm min}}
\left(
\frac{ e^{ - \frac{k_{\rm min}}{Nc} t } }{ 1 - \frac{c - k_{\rm min}}{c+1-k_{\rm min}} e^{ - \frac{1}{Nc} t } }
\right).
\label{eq:PTFRtExp}
\end{equation}

\noindent
To explore the asymptotic long time tail of $P(T_{\rm FR}>t)$ we expand
the right hand side of Eq. (\ref{eq:PTFRtExp}) in powers of 
$\exp \left( - \frac{t}{Nc} \right) \ll 1$. We obtain

\begin{equation}
P(T_{\rm FR} > t) \simeq
\frac{1}{c+1-k_{\rm min}}
e^{ - \frac{k_{\rm min}}{Nc} t }. 
\label{eq:PTFRtExp2}
\end{equation}

\noindent
As can be seen, this tail is dominated by the lowest degree nodes, 
whose degree is $k_{\rm min}$.

\begin{figure}
\centerline{
\includegraphics[width=9cm]{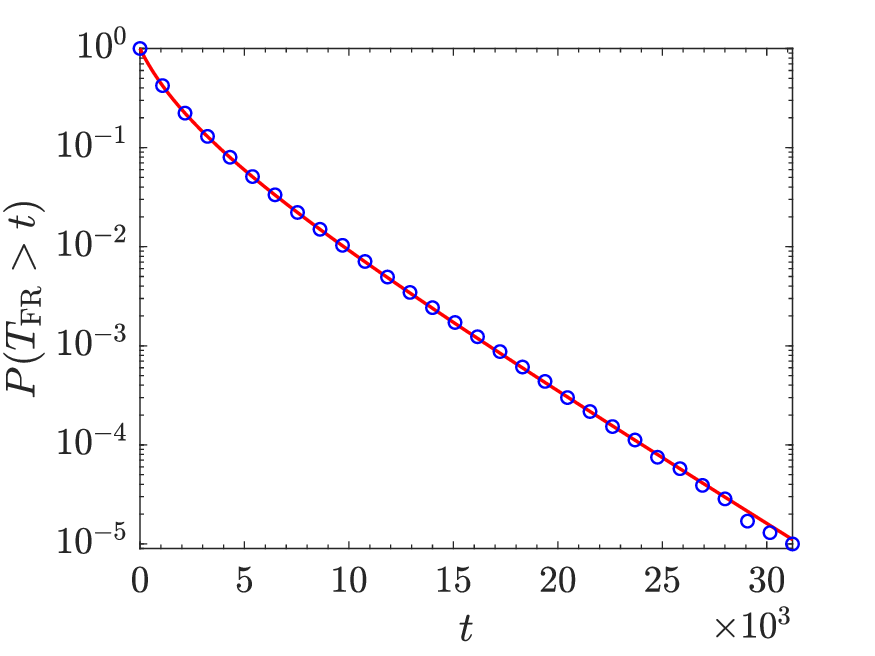}
}
\caption{
Analytical results for the 
tail distribution  
$P(T_{\rm FR} > t)$ (solid lines)
of first return times
of an NBW on a configuration model network
of size $N=1000$ 
which exhibits an exponential degree distribution with
$k_{\rm min}=3$ and
mean degree 
$c=10$.
The analytical results,
obtained from Eq.
(\ref{eq:PTFRtExp}),
are in very good agreement with the results
obtained from computer simulations (circles).
The simulation results were obtained using the same averaging procedure as in Fig. \ref{fig:1}.
}
\label{fig:5}
\end{figure}

In Fig. \ref{fig:5} we present
analytical results for the 
tail distribution  
$P(T_{\rm FR} > t)$ (solid lines)
of first return times
of an NBW on a configuration model network 
of size $N=1000$ 
that exhibits an exponential degree distribution
with $k_{\rm min}=3$
and mean degree 
$c=10$.
The analytical results,
obtained from Eq.
(\ref{eq:PTFRtExp}),
are in very good agreement with the results
obtained from computer simulations (circles).

To calculate the mean of the distribution of first return times, 
we use Eq. (\ref{eq:ETFR3}), and obtain

\begin{equation}
{\mathbb E} [ T_{\rm FR} ] =
N \frac{c}{c+1-k_{\rm min}} 
\Phi \left( \frac{c-k_{\rm min}}{c+1-k_{\rm min}},1,k_{\rm min} \right),
\label{eq:ETFR_Exp}
\end{equation}

\noindent
where $\Phi(z,s,\alpha)$ is the Lerch transcendent
\cite{Olver2010}.
In the limit of large mean degree $c$, we obtain

\begin{equation}
{\mathbb E} [ T_{\rm FR} ]   =
N \left[ \ln c - H_{k_{\rm min}-1} + {\mathcal O} \left( \frac{\ln c}{c} \right) \right],
\label{eq:ETFR_Exp2}
\end{equation}

\noindent
where $H_{m}$ is the Harmonic number
\cite{Olver2010}.
Note that the leading term in ${\mathbb E}[T_{\rm FR}]$ is given by $N \ln c$,
unlike the  RRG and ER network, in which ${\mathbb E}[T_{\rm FR}] \simeq N$.
This reflects the fact that the variance of the degrees in the exponential case is much
larger than in the Poisson distribution and that it increases as $c$ is increased.
Since the exponential distribution is broad and highly asymmetric,
the expansion presented in Eq. (\ref{eq:inverseK}) cannot be used to reproduce the results of
Eq. (\ref{eq:ETFR_Exp2}).
Eq. (\ref{eq:ETFR_Exp2}) shows that the mean first return time in configuration model
networks with an exponential degree distribution increases 
logarithmically with the mean degree $c$.

\begin{figure}
\centerline{
\includegraphics[width=9cm]{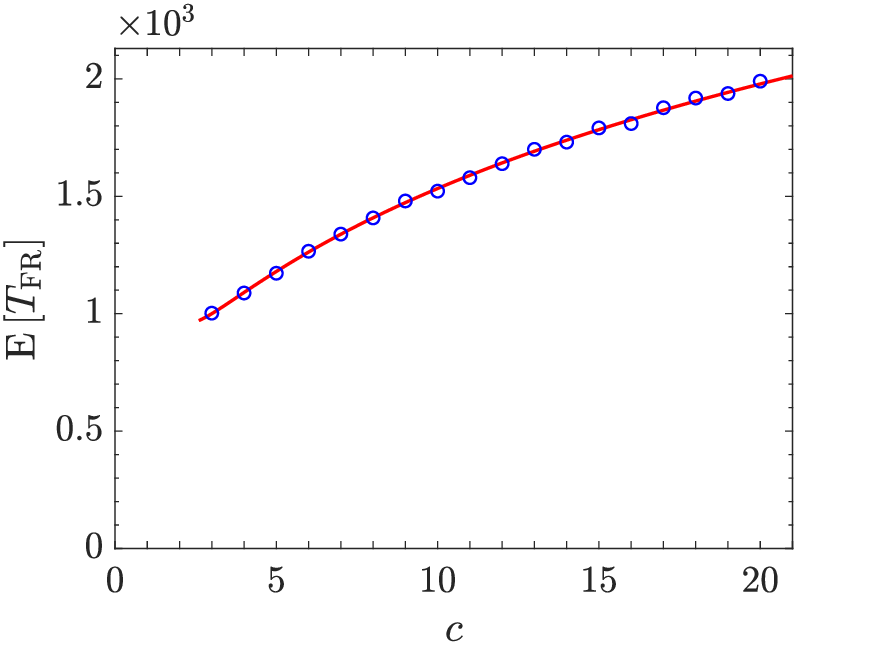}
}
\caption{
Analytical results for the
mean first return time 
${\mathbb E}[ T_{\rm FR} ]$ (solid line)
of an NBW on a configuration model network of size $N=1000$,
which exhibits an exponential degree distribution with $k_{\rm min}=3$,
as a function of the mean degree 
$c$.
The analytical results,
obtained from Eq.
(\ref{eq:ETFR_Exp}), 
are in very good agreement with 
the results obtained from computer simulations (circles).
The simulation results were obtained using the same averaging procedure as in Fig. \ref{fig:3}.
}
\label{fig:6}
\end{figure}

In Fig. \ref{fig:6} we present
analytical results for the
mean first return time 
${\mathbb E} [ T_{\rm FR}]$ (solid line)
of an NBW on a configuration model network of size $N=1000$
which exhibits an exponential degree distribution with $k_{\rm min}=3$,
as a function of the mean degree 
$c$.
The analytical results,
obtained from Eq.
(\ref{eq:ETFR_Exp}), 
are in very good agreement with 
the results obtained from computer simulations (circles).

Similarly, we calculate the second moment, using Eq. (\ref{eq:ET2FR}).
We obtain

\begin{equation}
{\mathbb E} \left[ T_{\rm FR}^2 \right ] =
2 N^2
\frac{c^2}{c+1-k_{\rm min}}
\Phi \left( \frac{c-k_{\rm min}}{c+1-k_{\rm min}},2,k_{\rm min} \right).
\end{equation}

\noindent
In the limit of large $c$, we obtain

\begin{equation}
{\mathbb E} \left[ T_{\rm FR}^2 \right ] =
2 N^2
\left[ (c+2 k_{\rm min} - 1) \zeta(2,k_{\rm min}) -   \ln c + 
H_{k_{\rm min}-1} - 1 + {\mathcal O} \left( \frac{ \ln c}{c} \right) \right],
\end{equation}

\noindent
where $\zeta(s,a)$ is the Hurwitz zeta function
\cite{Olver2010}.

The variance is given by

\begin{equation}
{\rm Var}(T_{\rm FR}) =
(Nc)^2
\left\{ \frac{ 2 \Phi \left( \frac{c-k_{\rm min}}{c+1-k_{\rm min}},2,k_{\rm min} \right) }{c+1-k_{\rm min}}
-
\left[
\frac{ \Phi  \left( \frac{c-k_{\rm min}}{c+1-k_{\rm min}},1,k_{\rm min} \right) }{ c+1-k_{\rm min}  } 
\right]^2
\right\}.
\label{eq:VarTFR_Exp}
\end{equation}

In the limit of large $c$, one can express the variance in a simpler form, namely

\begin{eqnarray}
{\rm Var}(T_{\rm FR}) & \simeq
N^2  \bigg[  2 c \  \zeta(2,k_{\rm min})  - (\ln c)^2 + 2 \left( H_{k_{\rm min}-1} - 1 \right) \ln c
\bigg.
\nonumber \\
& + 2 \left( 2 k_{\rm min} - 1 \right)  \zeta(2,k_{\rm min})
+ 2 H_{k_{\rm min} - 1} 
\nonumber \\
& - \left.  H_{k_{\rm min} - 1}^2 - 2
+ {\mathcal O} \left( \frac{ \ln c }{c} \right) \right].
\label{eq:VarTFR_Exp2}
\end{eqnarray}

\noindent
Note that the leading term is proportional to the mean degree $c$.

\begin{figure}
\centerline{
\includegraphics[width=9cm]{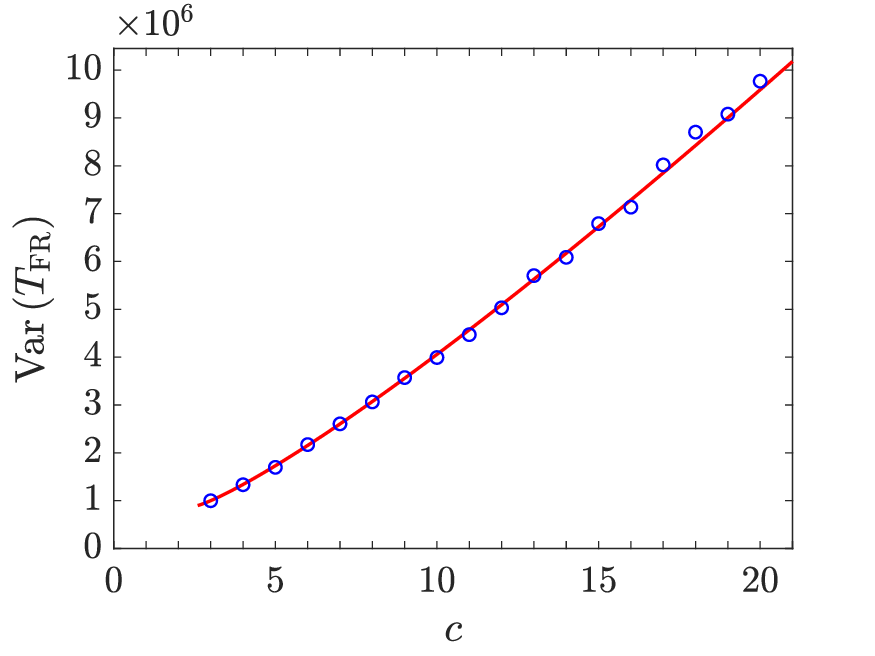}
}
\caption{
Analytical results for the
variance   
${\rm Var}(T_{\rm FR})$ (solid line)
of the distribution of first return times
of an NBW on a configuration model network of size $N=1000$,
which exhibits an exponential degree distribution with $k_{\rm min}=3$,
as a function of the mean degree 
$c$.
The analytical results,
obtained from Eq.
(\ref{eq:VarTFR_Exp}), 
are in very good agreement with 
the results obtained from computer simulations (circles).
The simulation results were obtained using the same averaging procedure as in Fig. \ref{fig:3}.
}
\label{fig:7}
\end{figure}

In Fig. \ref{fig:7} we present 
analytical results for the
variance   
${\rm Var}(T_{\rm FR})$ (solid line)
of the distribution of first return times
of an NBW on a configuration model network of size $N=1000$,
which exhibits an exponential degree distribution with $k_{\rm min}=3$,
as a function of the mean degree 
$c$.
The analytical results,
obtained from Eq.
(\ref{eq:VarTFR_Exp}), 
are in very good agreement with 
the results obtained from computer simulations (circles).

\subsection{Configuration model networks with a power-law degree distribution}
 
Consider a configuration model network with a power-law degree distribution of the form

\begin{equation}
P(k) = A k^{-\gamma},
\label{eq:Pk_SF}
\end{equation}

\noindent
where the degree $k$ takes values in the range
$k_{\rm min} \le k \le k_{\rm max}$.
The parameter
$A = [ \zeta(\gamma,k_{\rm min}) - \zeta(\gamma,k_{\rm max}+1) ]^{-1}$
is a normalization constant.
Here we focus on the case that $k_{\rm min} \ge 2$, in which the network 
consists of a single connected component.

Since a power-law distribution may allow nodes of high degree, 
it is important to note that in order to enable the construction of a configuration
model network in which degree-degree correlations are negligible,
one must impose an upper cutoff on the degree distribution,
which satisfies
$k_{\rm max} < \sqrt{Nc}$
\cite{Boguna2004,Catanzaro2005}.

The mean degree is given by
\cite{Tishby2018}

\begin{equation}
c = \langle K \rangle = 
\frac{ \zeta(\gamma-1,k_{\rm min}) - \zeta(\gamma-1,k_{\rm max}+1) }
{ \zeta(\gamma,k_{\rm min}) - \zeta(\gamma,k_{\rm max}+1) },
\label{eq:Kmsf}
\end{equation}

\noindent
and the second moment of the degree distribution is

\begin{equation}
\langle K^2 \rangle = 
\frac{ \zeta(\gamma-2,k_{\rm min}) - \zeta(\gamma-2,k_{\rm max}+1) }
{ \zeta(\gamma,k_{\rm min}) - \zeta(\gamma,k_{\rm max}+1) }.
\label{eq:K2msf}
\end{equation}

\noindent
The variance of the degree distribution is given by
${\rm Var}(K) = \langle K^2 \rangle - \langle K \rangle^2$.

For $\gamma \le 2$ the mean degree (and the variance) diverge when 
$k_{\rm max} \rightarrow \infty$.
For $\gamma > 3$ both the mean degree and the variance are bounded.
In the intermediate range of
$2 < \gamma < 3$ the mean degree $\langle K \rangle$ is 
bounded while the  
variance ${\rm Var}(K)$ diverges.
In this regime, 
as $k_{\rm max}$ is increased, the variance diverges like

\begin{equation}
{\rm Var}(K) \simeq
\frac{ 1 }
{(3-\gamma) [\zeta(\gamma,k_{\rm min}) - \zeta(\gamma,k_{\rm max}+1)] } 
(k_{\rm max})^{3-\gamma}.
\label{eq:VarKsf}
\end{equation}

Inserting $P(k)$ from Eq. (\ref{eq:Pk_SF}) into Eq. (\ref{eq:PTFRk3}) and carrying out the summation,
we obtain

\begin{equation}
P(T_{\rm FR}>t) =
\frac{ e^{ - \frac{t}{Nc} k_{\rm min} } \Phi \left( e^{ - \frac{t}{Nc} }, \gamma, k_{\rm min} \right)
-
e^{ - \frac{t}{Nc} (k_{\rm max} + 1)  } \Phi \left( e^{ - \frac{t}{Nc} }, \gamma, k_{\rm max} + 1 \right) }
{ \zeta(\gamma,k_{\rm min}) - \zeta(\gamma,k_{\rm max}+1) }.
\label{eq:PTFRtSF}
\end{equation}

\noindent
To explore the asymptotic long time tail of $P(T_{\rm FR}>t)$ we expand
the right hand side of Eq. (\ref{eq:PTFRtSF}) in powers of 
$\exp \left( - \frac{t}{Nc} \right) \ll 1$, we obtain

\begin{equation}
P(T_{\rm FR}>t) \simeq
\frac{ (k_{\rm min})^{- \gamma}  }
{ \zeta(\gamma,k_{\rm min}) - \zeta(\gamma,k_{\rm max}+1) }
e^{ - \frac{ k_{\rm min} }{Nc} t }.
\label{eq:PTFRtSF2}
\end{equation}

\noindent
As can be seen, this tail is dominated by the lowest degree nodes, 
whose degree is $k_{\rm min}$.

\begin{figure}
\centerline{
\includegraphics[width=9cm]{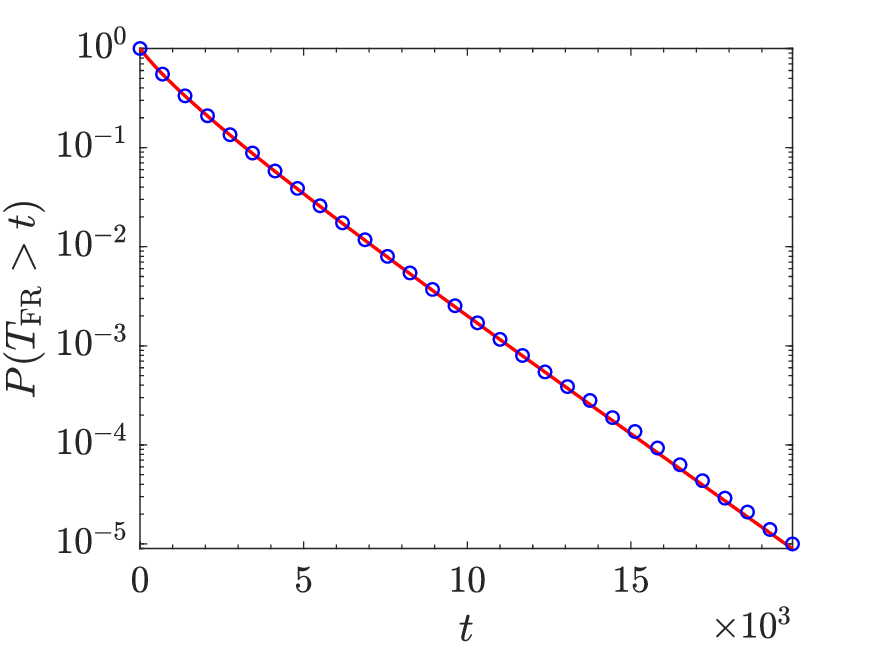}
}
\caption{
Analytical results for the 
tail distribution  
$P(T_{\rm FR} > t)$ (solid lines)
of first return times
of an NBW on a configuration model network
of size $N=1000$ 
which exhibits a power-law distribution with
$k_{\rm min}=3$, $k_{\rm max}=30$ and $\gamma=2.5$.
The analytical results,
obtained from Eq.
(\ref{eq:PTFRtSF}),
are in very good agreement with the results
obtained from computer simulations (circles).
The simulation results were obtained using the same averaging procedure as in Fig. \ref{fig:1}.
}
\label{fig:8}
\end{figure}

In Fig. \ref{fig:8} we present
analytical results for the 
tail distribution  
$P(T_{\rm FR} > t)$ (solid lines)
of first return times
of an NBW on a configuration model network 
of size $N=1000$ 
that exhibits a power-law degree distribution
with $k_{\rm min}=3$, $k_{\rm max} = 30$ and $\gamma = 2.5$,
which yields a mean degree of $\langle K \rangle \simeq 5.58$.
The analytical results,
obtained from Eq.
(\ref{eq:PTFRtSF}),
are in very good agreement with the results
obtained from computer simulations (circles).

Inserting $P(k)$ from Eq. (\ref{eq:Pk_SF}) into
Eq. (\ref{eq:ETFR3}) and carrying out the summation, 
we obtain the mean first return time,
which is given by

\begin{equation}
{\mathbb E} [ T_{\rm FR} ] =
N \langle K \rangle \frac{ \zeta(\gamma+1,k_{\rm min}) 
- \zeta(\gamma+1,k_{\rm max}+1) }
{ \zeta(\gamma,k_{\rm min}) - \zeta(\gamma,k_{\rm max}+1) }.
\label{eq:ETFR_PL}
\end{equation}

\noindent
Since the power-law distribution is broad and highly asymmetric,
the expansion presented in Eq. (\ref{eq:inverseK}) 
for $\langle \frac{1}{K} \rangle$
cannot be used to reproduce the results of
Eq. (\ref{eq:ETFR_PL}).

In the limit of $k_{\rm max} \rightarrow \infty$, 
inserting $\langle K \rangle$ from Eq. (\ref{eq:Kmsf}),
Eq. (\ref{eq:ETFR_PL}) is reduced to

\begin{equation}
{\mathbb E} [ T_{\rm FR} ] =
N   \frac{ \zeta(\gamma-1,k_{\rm min})  \zeta(\gamma+1,k_{\rm min}) }
{ [\zeta(\gamma,k_{\rm min})]^2   }.
\label{eq:ETFR_PL2}
\end{equation}
 
\noindent
For $\gamma \ne 1$ and $k > 0$, the Hurwitz zeta function can be expressed in the form

\begin{equation}
\zeta(\gamma,k) = \frac{ k^{- \gamma}}{2} + \frac{k^{1-\gamma}}{\gamma-1}
+ \frac{1}{\Gamma(\gamma)} \int_{0}^{\infty} 
\left( \frac{1}{e^x - 1} - \frac{1}{x} + \frac{1}{2} \right) x^{\gamma-1} e^{- k x} dx,
\label{eq:zeta1}
\end{equation}

\noindent
where $\Gamma(y)$ is the Gamma function
\cite{Olver2010}.
In the context of this paper the Hurwitz zeta function $\zeta(\gamma,k)$ is evaluated in the range of
$\gamma > 1$ and $k \ge 3$. 
Exploring the terms on the right hand side of Eq. (\ref{eq:zeta1}) in this range of values,
it was found that the contribution of the integral is negligible.
Thus, Eq. (\ref{eq:zeta1}) can be simplified to

\begin{equation}
\zeta(\gamma,k) \simeq \frac{1}{2} k^{- \gamma}
\left( 1 + \frac{2 k}{\gamma - 1} \right).
\label{eq:zeta2}
\end{equation}

\noindent
Inserting $\zeta(\gamma,k)$ from Eq. (\ref{eq:zeta2}) into Eq. (\ref{eq:ETFR_PL2}),
we obtain

\begin{equation}
{\mathbb E} [ T_{\rm FR} ] \simeq
N    \frac{ \left( 1 + \frac{2 }{\gamma-2} k_{\rm min} \right)\left( 1 + \frac{2 }{\gamma} k_{\rm min} \right) }
{ \left( 1 + \frac{2 }{\gamma-1} k_{\rm min} \right)^2 }.
\label{eq:ETFR_PL2p}
\end{equation}

\noindent
For sufficiently large values of $k_{\rm min}$, 
Eq. (\ref{eq:ETFR_PL2}) can be approximated by

\begin{equation}
{\mathbb E} [ T_{\rm FR} ] \simeq
N  \left[    \frac{ (\gamma-1)^2 }{ \gamma (\gamma-2) }  + {\mathcal O} \left( \frac{1}{k_{\rm min}} \right) \right].
\label{eq:ETFR_PL3}
\end{equation}

\noindent
In practice, for $k_{\rm min} = 3$ there is a slight deviation between 
the right hand sides of Eqs. (\ref{eq:ETFR_PL2}) and (\ref{eq:ETFR_PL3}),
which becomes negligible for $k_{\rm min} \ge 5$.
It is found that ${\mathbb E} [ T_{\rm FR} ]$ is a monotonically
decreasing function of the exponent $\gamma$.
In the limit of $\gamma \gg 1$ it converges towards
${\mathbb E} [ T_{\rm FR} ] \simeq N$,
where it coincides with the result for RRGs.
In the opposite limit, when $\gamma \rightarrow 2^{+}$ the mean first
return time diverges (given that $k_{\rm max} \rightarrow \infty$).

Going back to Eq. (\ref{eq:ETFR_PL}), taking the limit of $k_{\rm max} \gg k_{\rm min}$
and using Eq. (\ref{eq:zeta2}) to approximate the ratio 
$\zeta(\gamma+1,k_{\rm min})/\zeta(\gamma,k_{\rm min})$
while leaving $\langle K \rangle$ unchanged, we obtain

\begin{equation}
{\mathbb E} [ T_{\rm FR} ] \simeq
N \frac{\gamma - 1}{ \gamma } \frac{ \langle K \rangle }{ k_{\rm min} }.
\label{eq:ETFR_PL4}
\end{equation}

\noindent
While the results obtained from Eq. (\ref{eq:ETFR_PL4}) are not as 
accurate as those obtained from Eq. (\ref{eq:ETFR_PL3}),
it provides useful insight on the relation between the mean first return time
and the mean degree $\langle K \rangle$.
Comparing Eq. (\ref{eq:ETFR_PL4}) to Eq. (\ref{eq:ETFR4})
shows that the mean first return time is dominated by the lowest degree nodes.

The second moment, obtained from Eq. (\ref{eq:ET2FR}),
is given by

\begin{equation}
{\mathbb E} \left[ T_{\rm FR}^2 \right]  = 
2 (N \langle K \rangle)^2 \frac{ \zeta(\gamma+2,k_{\rm min}) - \zeta(\gamma+2,k_{\rm max}+1) }
{ \zeta(\gamma,k_{\rm min}) - \zeta(\gamma,k_{\rm max}+1) },
\end{equation}

\noindent
and the variance is given by

\begin{eqnarray}
{\rm Var}(T_{\rm FR}) &=&
2 (N \langle K \rangle)^2  \frac{ \zeta(\gamma+2,k_{\rm min}) - \zeta(\gamma+2,k_{\rm max}+1) }
{ \zeta(\gamma,k_{\rm min}) - \zeta(\gamma,k_{\rm max}+1) }
\nonumber \\
&-&  (N \langle K \rangle)^2 
\left[ \frac{ \zeta(\gamma+1,k_{\rm min}) - \zeta(\gamma+1,k_{\rm max}+1) }
{ \zeta(\gamma,k_{\rm min}) - \zeta(\gamma,k_{\rm max}+1) } \right]^2.
\label{eq:VarTFR_sf1}
\end{eqnarray}

\noindent
In the limit of $k_{\rm max} \rightarrow \infty$, 
and for values of $\gamma$ which are sufficiently far above $\gamma=2$,  the 
variance of the distribution of first return times
can be roughly approximated by

\begin{eqnarray}
{\rm Var} ( T_{\rm FR} ) &\simeq&   
2 N^2  
\frac{ \left( 1 + \frac{2 }{\gamma-1} k_{\rm min} \right)^2 \left( 1 + \frac{2 }{\gamma+2} k_{\rm min} \right) }
{ \left( 1 + \frac{2 }{\gamma} k_{\rm min} \right)^3 }
\nonumber \\
&-& N^2 
\frac{ \left( 1 + \frac{2 }{\gamma-1} k_{\rm min} \right)^2 \left( 1 + \frac{2 }{\gamma+1} k_{\rm min} \right)^2 }
{ \left( 1 + \frac{2 }{\gamma} k_{\rm min} \right)^4 }.
\label{eq:VarTFR_sf2}
\end{eqnarray}

\noindent
For sufficiently large values of $k_{\rm min}$, Eq. (\ref{eq:VarTFR_sf2}) can be approximated by

\begin{equation}
{\rm Var} ( T_{\rm FR} )  \simeq   N^2
\frac{ \gamma^3 ( \gamma^2 + 2 \gamma + 2) }{(\gamma-1)^2 (\gamma+1)^2 (\gamma+2) }.
\label{eq:VarTFR_sf3}
\end{equation}

\noindent
In the limit of large $\gamma$, the variance converges towards $N^2$, in agreement
with the result for RRGs.
Interestingly, in the opposite limit of
$\gamma \rightarrow 2^{+}$
the variance remains finite,
unlike the mean first return time that tends to diverge.

\section{Discussion}

A key observation, expressed by Eq. (\ref{eq:PTFRk2}) is that the distribution of first return times
for an initial node of a given degree $k$ depends only on the degree $k$ and on the total number
of 'directed' edges in the network, given by $Nc$.
It does not depend on the degree distribution $P(k)$, which accounts for the way in which the
$Nc-2k$ remaining 'directed' edges are divided among the $N-1$ remaining nodes.
This implies that the distribution of first return times is determined by local properties
of the network and is not sensitive to the global structure.
 
\begin{figure}
\centerline{
\includegraphics[width=9cm]{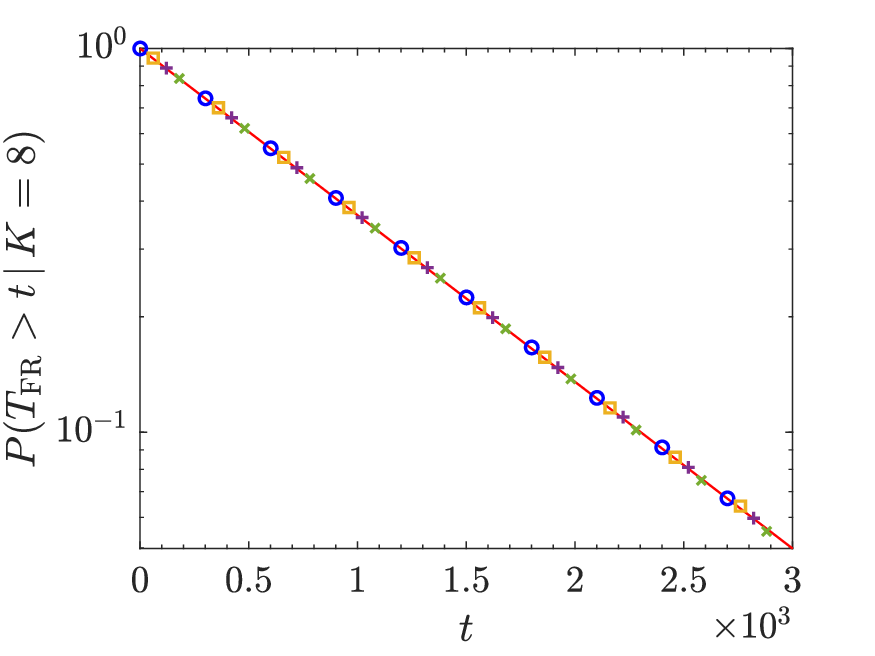}
}
\caption{
Analytical results for the distribution of first return times
for NBWs starting from an initial node of degree $k=8$ in configuration model
networks of size $N=1000$ and mean degree $c=8$, given by Eq. (\ref{eq:PTFRk2}).
The analytical results are found to be in very good agreement with simulation
results for RRGs ($\circ$), ER networks ($\times$) and configuration model networks
with exponential ($+$) and power-law ($\square$) distributions.
Each data point in the simulation results was obtained by averaging over
$20$ independent network instances and $10,000$ NBW trajectories
in each network instance, starting from the same initial node of degree $k=8$.
}
\label{fig:9}
\end{figure}

In Fig. \ref{fig:9} we present analytical results for the distribution of first return times
of NBWs starting from a random initial node of degree $k=8$ in a configuration model
network of size $N=1000$ and mean degree $c=8$, given by Eq. (\ref{eq:PTFRk2}).
The analytical results are found to be in very good agreement with simulation
results for RRGs ($\circ$), ER networks ($\times$) and configuration model networks
with exponential ($+$) and power-law ($\square$) distributions.
As can be seen, the distribution $P(T_{\rm FR} > t | K=8)$ does not depend
on the degree distribution $P(k)$ but only on the mean degree $c$ 
and on the degree of the initial node.

Comparing the results obtained for the four random network models considered above,
we conclude that the mean first return time strongly depends on the variability of the
degrees of nodes in the network. More specifically, as the degree distribution $P(k)$ becomes
broader the mean first return time ${\mathbb E}[T_{\rm FR}]$ increases.
This is illustrated by the fact that for an NBW on an RRG 
${\mathbb E}[T_{\rm FR}] \simeq N$,
for an NBW on an ER network
${\mathbb E}[T_{\rm FR} | K>1] \simeq N \left( 1 + \frac{1}{c} \right)$,
for an NBW on a configuration model network with an exponential degree distribution
${\mathbb E}[T_{\rm FR}] \simeq N \ln c$
and for an NBW on a configuration model network with a power-law degree distribution
${\mathbb E}[T_{\rm FR}] \sim Nc/k_{\rm min}$.
In light of these results, it is interesting to note that the dependence 
of the mean first return time on the mean degree $c$
is a non-trivial issue, which depends on the details of the degree distribution.
In the examples studied here we observe three different behaviors:
in RRGs the mean first return time is independent of $c$,
in ER networks it decreases with $c$ and in configuration model networks with
an exponential degree distribution it increases with $c$.

In all the network ensembles considered above, the long-time tail of $P(T_{\rm FR}>t)$
exhibits a decaying exponential form, which is determined by the lowest-degree nodes
in the network. From a broader perspective, it implies that the distribution of first return 
times is mostly characterised by low-degree nodes that reside in the periphery of the network.
This is unlike the outburst dynamics of
other processes such as the spreading of information and infections, 
which are dominated by the highest degree nodes (or hubs) that reside in the core of the network
\cite{Barrat2012}.

Apart from the first return process, there are other significant events that take place
over the lifetime of an NBW (and other RWs) on a random network. 
One of them is the first hitting (FH) process,
which is the first time at which an NBW steps into a previously visited node. 
Starting from a random initial node $i$, in the early stages of its trajectory, an NBW visits
a new node at each time step. 
During this time, the statistical properties of the NBW trajectory are identical to those
of a self avoiding walk 
\cite{Tishby2016}.
After the first hitting event, in some of the time steps the NBW visits yet-unvisited nodes and in
other time steps it revisits nodes that it has already visited before.
The distribution of first hitting times of RWs and NBWs on ER networks were studied in Refs.
\cite{Tishby2017} and \cite{Tishby2017b}, respectively.
It was found that in both cases, for sufficiently dense ER networks in which there are no leaf nodes
of degree $k=1$, the distribution 
$P(T_{\rm FH}>t)$
of first hitting times
is given by a product of an exponential distribution and
a Rayleigh distribution, which is a special case of the Weibull distribution.
In this limit, the mean first hitting time of NBWs on ER networks is given by

\begin{equation}
{\mathbb E}[T_{\rm FH}] = \sqrt{ \frac{\pi}{2} } \sqrt{N}.
\end{equation}

\noindent
Similar results were also obtained for first hitting processes on RRGs
\cite{Tishby2021b}.

The results presented in this paper shed light on the more general class of first passage processes.
Consider an NBW starting from a random initial node $i$, seeking a target node $j$, where $j \ne i$.
Unlike the first return event of an NBW which may take place only at $t \ge 3$, a first passage
event may take place even at $t=1$ (in case that $i$ and $j$ are connected by an edge).
We thus conclude that to a very good approximation, the distribution of first passage times of
NBWs on configuration model networks can be expressed in the form

\begin{equation}
P(T_{\rm FP}>t) \simeq P(T_{\rm FR}>t+2).
\end{equation}

Another important event, which occurs at much longer time scales, is the step at
which an NBW (or RW) completes visiting all the nodes in the network. The time at which this
happens is called the cover-time.
For RWs on RRGs it was shown that the mean cover time scales like

\begin{equation}
{\mathbb E}[T_{\rm C}]  \propto N \ln N.
\label{eq:CT}
\end{equation}

\noindent
This means that
on average an RW visits each node $\ln N$ times before it completes visiting all the nodes
in the network at least once.
The distribution of cover times of RWs on RRGs was studied in Refs.
\cite{Cooper2005,Masuda2017,Tishby2022a}.
Since they do not backtrack their steps, NBWs scan the network more efficiently than RWs.
This is expected to affect the pre-factor of the scaling relation 
on the right hand side of Eq. (\ref{eq:CT})
but is not expected to change the way the cover time scales with $N$.

The results presented in this paper were derived in the context of configuration model networks.
However, we expect them to apply within a good approximation to a somewhat broader class of 
small-world networks which are sufficiently strongly connected without bottlenecks 
and exhibit short mixing times
of the NBW, determined by the spectral gap of the 
non-backtracking (Hashimoto) matrix.
In contrast, these results are not expected to apply in the case of modular networks,
which consist of several modules with weak connections between them and for networks
that exhibit long mixing times.
However, in the case of modular networks, 
at short times, the distribution of first return times is likely to 
behave as if the module on which the initial node resides
is isolated from other modules.
This behavior persists until the probability that the NBW will hop into
some other module becomes significant.

In essence, the derivation presented above requires that the mixing time will be much shorter
than the mean first return time, such that the assumption that the NBW samples uniformly
the 'directed' edges is justified.
In light of this it is interesting to discuss the effect of degree-degree correlations.
In general, negative or disassortative correlations tend to enhance the connectivity
of the network
\cite{Tishby2018}
and hence shortens the mixing times
\cite{Noldus2015}
by increasing the spectral gap.
On the other hand, positive or assortative correlations are known to 
decrease the spectral gap
\cite{Mieghem2010},
thus increasing the mixing time.
This is particularly relevant in networks that have many high degree nodes,
such as scale-free networks where high assortativity may break the network 
into disconnected components
\cite{Tishby2018}.
However, for low correlations the overall impact of degree-degree correlations on
the spectrum is not large,
especially on short range correlations between eigenvalues that
follow the predictions of random matrix theory
\cite{Jalan2015}.
In summary, in the case of disassortative networks we expect our results to hold.
Regarding networks that exhibit low to mild positive assortativity and to the extent that
they do not break the network into disconnected components, we expect the
results to hold to a good approximation.

Another key factor influencing the mixing times is the clustering coefficient, primarily
mediated through its effect on the spectral gap.
The clustering coefficient measures how often a node's neighbors form triangles
indicating the degree of local inter-connectedness.
Networks with higher clustering coefficients typically have smaller
spectral gaps. This occurs because increased clustering introduces more
local structure.
A smaller spectral gap leads to longer mixing times, since random walks become
trapped in tightly connected neighborhood before fully exploring the network
\cite{Minh2024}.
Consequently, higher clustering increases the mixing times.
We thus expect our results to be valid as long as the clustering is not too strong.

While directed networks are of significant theoretical interest, they introduce complexities
that fall outside the scope of this paper.
In directed networks the asymmetry of edges creates distinct behavior as the random walk
dynamics are heavily influenced by both the in-degrees and out-degrees.
This asymmetry complicates the analysis of return times and of the mixing behavior,
often leading to nodes with low in-degrees being visited only rarely or potentially 
not at all.
Additionally, in weakly connected networks random walkers may become trapped in
certain domains rendering the analysis of return times more intricate.
As the focus of this paper is on undirected networks, we leave these issues to future work.

\section{Summary}
 
 We presented analytical results for the distribution of
first return times of
NBWs on configuration model networks
consisting of $N$ nodes with degree distribution $P(k)$,
focusing on the case in which the network consists of a single connected component.
It was found that the tail distribution
$P ( T_{\rm FR} > t )$
of first return times 
is given by a discrete Laplace transform of the degree distribution $P(k)$.
This result demonstrates the relation between structural properties of a network,
captured by the degree distribution,
and the properties of dynamical processes taking place on the network.
It was found that $P( T_{\rm FR} > t )$ exhibits an exponential tail,
which is determined by the properties of the low-degree nodes that reside
in the periphery of the network.
We calculated the mean first return time and found that
${\mathbb E}[ T_{\rm FR} ] = \langle \frac{Nc}{K} \rangle$. 
Surprisingly, this result
coincides with the result of Kac's lemma that applies to simple RWs,
in agreement with recent rigorous results by Fasino et al.
\cite{Fasino2023}.
We also calculated the variance ${\rm Var}(T_{\rm FR})$, which accounts
for the variability of the first return times between different NBW trajectories.
We applied this formalism to random regular graphs, Erd{\H o}s-R\'enyi
networks and configuration model networks with exponential and
power-law degree distributions and obtained closed-form expressions for
$P ( T_{\rm FR} > t )$ and its first two moments.
These results provide useful insight on the advantages of NBWs over
simple RWs in network exploration, sampling and search processes.
Our results are expected to hold for a broader class of networks, in
which the mixing time is much shorter than the mean first return times.

This work was supported by Grant no. 2020720 from the 
United States-Israel Binational Science Foundation (BSF)
and grant no. 2102832 from the National Science Foundation (NSF).

\end{document}